\documentclass[11pt,a4paper]{article}
\usepackage{jheppub} % for details on the use of the package, please
\newcommand{\Snhat}{\hat{\mathbf{S}}_n}

\newcommand{\mre}{\mathrm{e}}
\newcommand{\mrO}{\mathrm{O}}
\newcommand{\mrd}{\mathrm{d}}

\newcommand{\bS}{\mathbf{S}}

\newcommand{\Z}{{\mathbb{Z}}}
\newcommand{\R}{{\mathbb{R}}}
\newcommand{\IDR}{\overline{I}}

\newcommand{\pmu}{\partial_\mu}
\newcommand{\pnu}{\partial_\nu}
\newcommand{\mcR}{\mathcal{R}}

\newcommand{\mcWov}{\overline{\mathcal{W}}}
\newcommand{\Wov}{\overline{W}}
\newcommand{\cov}{\overline{c}}
\newcommand{\vpi}{\vec{\pi}} 
\newcommand{\lr}[1]{l_{#1}^\mathrm{r}}
\newcommand{\psump}{\sum_{p}\rule{0pt}{2.5ex}'\;}
\newcommand{\order}[1]{\mathrm{O}\left( #1 \right) }

\newcommand{\bfp}{\mathbf{p}}
\newcommand{\bfx}{\mathbf{x}}
\newcommand{\bR}{\mathbf{R}}

\newcommand{\mcVD}{\mathcal{V}_D}
\newcommand{\VdD}{\overline{V}_D}

\newcommand{\Nf}{N_\mathrm{f}}

\newcommand{\Ls}{L_s}
\newcommand{\Lt}{L_t}
\newcommand{\Lhat}{\widehat{L}}
\newcommand{\ellhat}{\hat{\ell}}

\DeclareMathOperator{\Tr}{Tr}
\DeclareMathOperator{\Res}{Res}
\DeclareMathOperator{\floor}{floor}
\newcommand{\fl}[1]{\left\lfloor {#1} \right\rfloor}

\title{\boldmath Isospin susceptibility in the O($n$) sigma-model 
  in the delta-regime}

\author[a]{F.\ Niedermayer} \author[b]{and P.\ Weisz}

\affiliation[a]{Albert Einstein Center for Fundamental Physics, \\
  Institute for Theoretical Physics, University of Bern, 
  Switzerland} 
\affiliation[b]{Max-Planck-Institut f\"ur Physik, 80805
  Munich, Germany}

\emailAdd{niedermayer@itp.unibe.ch} \emailAdd{pew@mpp.mpg.de}

\abstract{We compute the isospin susceptibility
  in an effective O($n$) scalar field theory (in $d=4$ dimensions),
  to third order in chiral perturbation theory ($\chi$PT) 
  in the delta--regime using the quantum mechanical rotator picture. 
  This is done in the presence of an additional coupling, 
  involving a parameter $\eta$, describing the effect of 
  a small explicit symmetry breaking term (quark mass).
  For the chiral limit $\eta=0$ we
  demonstrate consistency with our previous $\chi$PT computations 
  of the finite-volume mass gap and isospin susceptibility. 
  For the massive case by computing the leading mass effect in the 
  susceptibility using $\chi$PT with dimensional regularization,
  we determine the $\chi$PT expansion for $\eta$ to third order.
  The behavior of the shape coefficients for long tube geometry
  obtained here might be of broader interest.
  The susceptibility calculated from the rotator approximation differs
  from the $\chi$PT result in terms vanishing like $1/\ell$ for
  $\ell=L_t/L_s\to\infty$.
  We show that this deviation can be described by a correction to the
  rotator spectrum proportional to the square of the quadratic 
  Casimir invariant.}

\subheader{\hfill MPP-2017-36}

\begin{document}

\maketitle

\section{Introduction}
\label{Intro}

In a previous paper \cite{Niedermayer:2016yll} we computed 
the change in the free energy due to a chemical potential
coupled to a conserved charge in the non-linear O($n$) sigma model 
with two regularizations, lattice regularization (with standard action) 
and dimensional regularization (DR) in a general $d$--dimensional asymmetric
$\Ls^{d-1}\times\Lt$ volume
with periodic boundary conditions in all directions. 
This allowed us, for $d=4$, to establish two
independent relations among the 4--derivative couplings appearing 
in the effective Langrangians and in turn this enables
conversion of results for physical quantities computed by the lattice
regularization to those involving scales introduced in DR.  

In particular we could convert the computation of the mass gap
in a periodic box, by Niedermayer and Weiermann \cite{Niedermayer:2010mx} 
using lattice regularization to a result involving parameters of the
dimensionally regularized effective theory, and we verified this
result by a direct computation \cite{Niedermayer:2016yll}.
The proposal to measure the low-lying stable masses 
in the delta--regime to constrain some of
the low energy constants in the effective chiral Langrangian
describing low energy pion dynamics of QCD, 
was first made by Hasenfratz \cite{Hasenfratz:2009mp}.
For two flavors of massless quarks, the relevant $\chi$PT has
SU(2)$\times$SU(2)$\simeq$O(4) symmetry. 

The so called $\delta$--regime referred to above, 
is where the system is in a periodic spatial box of sides $\Ls$
with $\Lt \gg \Ls$
and $m_\pi\Ls$ is small (i.e. small or zero quark mass) 
whereas $F_\pi\Ls$, ($F_\pi$ the pion decay constant) is large. 
This regime was first analyzed in a pioneering 
paper by Leutwyler \cite{Leutwyler:1987ak}.
He showed that the low lying dynamics in this regime was
described by a quantum rotator for the spatially
constant modes. 

In that paper Leutwyler also obtained the leading order
effects of the quark mass on the low-lying spectrum for O(4).
The extension to next-to-leading (NLO) order was presented 
by Weingart \cite{Weingart:2010yv,Weingart:2010zz}. 
There followed a period of low activity in this field,
but recently Matzelle and Tiburzi have,
in an interesting paper \cite{Matzelle:2015lqk},
studied the effect of small symmetry breaking in the QM rotator 
picture ($\Nf=2$), and extended the results to small non-zero 
temperatures.

The rotator Hamiltonian given in \eqref{HMh} has
two parameters, the moment of inertia $\Theta$ and 
a parameter $\eta$ describing 
the explicit symmetry breaking O($n$) to O($n-1$). 
These parameters are themselves functions of $F,\Ls,M$ where $F$ the
pion decay constant in the limit of vanishing quark mass,
and $M$ is the mass acquired by the Goldstone bosons 
in leading order $\chi$PT
due to the explicit symmetry breaking.

Within this theoretical framework, in this paper 
we communicate computations which provide, in our opinion, 
a non-trivial check
on our previous rather technical computations \cite{Niedermayer:2016yll},
and also gives the NNLO (next-to next-to leading order)
$\chi$PT expansion for the parameter $\eta$.

In sect.~\ref{F_rot} we obtain the expression \eqref{chi_rotm} 
for the susceptibility for $u\equiv\Lt/(2\Theta)$ small,
under the assumption that in the $\delta$--regime
the low lying spectrum is that of the quantum rotator.
These computations are themselves rather involved and so 
we defer many details to Appendix~\ref{AppRot}. 
There we also review the leading order effects of a small 
symmetry breaking parameter on the low lying spectrum for general $n$, 
reproducing Leutwyler's result \cite{Leutwyler:1987ak} for $n=4$.

In sub-section~\ref{symmcase} we first consider the symmetric case $M=0=\eta$.
By inserting our expression for $\Theta$ obtained
in \cite{Niedermayer:2016yll} into \eqref{chi_rot0},
we obtain the susceptibility
for small $u$ as a function of $F,\Ls,\ell\equiv\Lt/\Ls$.
This can then be compared 
to the direct $\chi$PT computation of the susceptibility
\cite{Niedermayer:2016yll} in the $\epsilon$--regime for $\ell \gg 1$.
The comparison requires knowledge of the large $\ell$--behavior
of shape functions and the sunset integral appearing in the latter;
these are discussed in Appendix~\ref{AppShC}.
The agreement of the two computations provides  
the non-trivial check referred to above. 

In sub-section~\ref{PTmass} we compute the susceptibility
for small quark masses in the framework of $\chi$PT.  
Subsequently a comparison to the result \eqref{chi_rotm}
from the rotator computation determines the term in
$\eta$ proportional to $M^2$ as a function of $F\Ls$ to NNLO. 
Various technical aspects are deferred 
to Appendices~\ref{Apppertdetails} and \ref{AppHaar}.  

In section~\ref{Distortion} we investigate 
how the isospin susceptibility calculated in the $\delta$-regime 
of $\chi$PT approaches for increasing $\ell$
the result of the rotator approximation.
This turns out to be $\sim (F L_s)^{-4} \ell^{-1}$, 
and we determine the distortion of the standard rotator spectrum 
needed to describe this deviation.

\section{\boldmath The free energy of the O($n$) 
  rotator with isospin chemical potential}
\label{F_rot}

We first consider the case with unbroken symmetry and subsequently 
compute the leading order effect due to the presence of 
a small O($n$) symmetry breaking mass term.  

\subsection{Symmetric case}
\label{Rot0}

The Hamiltonian of the O($n$) quantum rotator with a chemical potential 
coupled to the generator $L_{12}$ of rotations in the $12$--plane is
\begin{equation}
  H_0(h) = \frac{\hat{L}^2}{2\Theta} - h \hat{L}_{12} \,,
  \label{H0h} 
\end{equation}
where $\Theta$ is the moment of inertia; to lowest order $\chi$PT
one has
$\Theta\simeq F^2\Ls^3$.
The energy levels and the corresponding multiplicities $g^{(n)}_{lm}$ 
are found in App.~\ref{AppRot}. 

The corresponding partition function for Euclidean time extent $\Lt$
and its expansion to $\order{h^2}$ is given by
\footnote{The notation $l$ is used below because of the 
analogy with the rotator, here it stands for the isospin.}
\begin{equation}
  \begin{aligned}
    Z(h;\Theta) &= \sum_{l=0}^\infty \sum_{m=-l}^l g^{(n)}_{lm} 
    \exp\left\{ -\left( \frac{\mathcal{C}_{n;l}}{2\Theta}-hm\right)\Lt\right\}
    \\
    &= z_0(u) + \frac12 h^2 \Lt^2 z_1(u) + \mrO(h^4)\,,
    \label{Z0h} 
  \end{aligned}
\end{equation}
where $u=\Lt/(2\Theta)$ and 
\begin{equation}
\mathcal{C}_{n;l}=l(l+n-2)\,,
\end{equation}
is the eigenvalue of the quadratic Casimir for isospin $l$.

The  coefficients $z_0(u)$, $z_1(u)$ and their expansions for small $u$ 
are given in App.~\ref{AppRot}. Using these results,
the isospin susceptibility of the O($n$) rotator for the unbroken case
is given by 
\begin{equation}
  \begin{aligned}
    \chi &= \frac{1}{\Lt V_s}\left. 
      \frac{\partial^2}{\partial h^2}\ln Z(h;\Theta)\right|_{h=0} 
    \\
    &= \frac{2\Theta}{n\Ls^3}\left[1-\frac{\Lt}{6\Theta}(n-2)
      +\frac{\Lt^2}{180 \Theta^2}(n-2)(n-4) + \ldots \right]\,.
  \end{aligned}
  \label{chi_rot0}
\end{equation}

\subsection{Rotator in external field}
\label{RotM}

Consider the effective Lagrangian with a term, 
breaking the O($n$) symmetry down to O($n-1$).
In the delta-regime this corresponds to an effective O($n$)
rotator in an external ``magnetic field'',
given by the Hamiltonian (including the isospin chemical potential)
\begin{equation}
  H(h) = H_0(h) + \eta \Snhat =
  \frac{\hat{L}^2}{2\Theta} + \eta \Snhat - h \hat{L}_{12} \,.
  \label{HMh} 
\end{equation}
Here $\Snhat$ is the $n$-th component of the operator defined as
$\hat{\mathbf{S}}\psi(\mathbf{S}) = \mathbf{S}\psi(\mathbf{S})$,
where $\mathbf{S}^2=1$. Its matrix elements between energy eigenstates 
of the Hamiltonian with $\eta=0$ are given 
in eq.~\eqref{v_lk}.

In the effective theory to leading order 
$\eta \sim M^2 \Theta \sim M^2 F^2 \Ls^3$.
The spectrum and the multiplicities are described in \ref{AppRot12}

Expanding the partition function in $h$ and $\eta$ to NL order
one obtains (cf.~\ref{AppRot22})
\begin{equation}
  \begin{aligned}
    Z(h;\Theta,\eta) &= \sum_{l=0}^\infty\sum_{k=0}^l\sum_{m=-k}^k g^{(n-1)}_{km} 
    \exp\left\{ -\left( E^{(n)}(l,k)-hm \right) \Lt \right\}
    \\
    &= z_0(u) - \eta^2\Theta \Lt  z_2(u) + \frac12 h^2 \Lt^2 
    \left[ z_1(u) - \eta^2 \Theta \Lt z_3(u)\right] + \ldots
    \label{ZMh} 
  \end{aligned}
\end{equation}
The expansion of $z_i(u)$ for small $u$ is given in Appendix~\ref{AppRot}.
The final result for the susceptibility of the rotator in an external 
magnetic field is 
\begin{equation}
  \begin{aligned}
    \chi &=  \frac{2\Theta}{n V_s}
    \Bigg\{ 1 - \frac13 (n-2)u + \frac{1}{45}(n-2)(n-4) u^2
    \\
    & \qquad \qquad
    - \frac{\eta^2 \Lt^2}{n(n+2)} \left[ 1 - \frac16 (2n-5)u
    +\frac{1}{60} (n^2-12n+17)u^2  \right]
    +\ldots \Bigg\}\,.
  \end{aligned}
  \label{chi_rotm}
\end{equation}

\section{Perturbative computation of the free energy}

The expression \eqref{chi_rotm} for the susceptibility 
was obtained under the assumption that the low lying 
spectrum in the $\delta$--regime is that of the quantum rotator.
In the next sub-section we first consider the symmetric case
and show the consistency of \eqref{chi_rot0} with our
previous computation of the susceptibility in \cite{Niedermayer:2016yll}
in the $\epsilon$--expansion.

In the $\epsilon$--expansion one considers a 4d box of fixed shape, 
in the case of an $\Ls^3\times\Lt$ volume a fixed aspect ratio $\ell=\Lt/\Ls$.
After separating the constant mode (the direction of the average magnetization
over the 4d volume), integration over the $p\ne 0$ modes result in 
an expansion in powers of $(F^2 L^2)^{-1}$, where by convention 
$L=(\Ls^3 \Lt)^{1/4}$. (For simplicity, 
here we discuss only the case without explicit symmetry breaking.)
The coefficients of the expansion contain the shape coefficients 
depending on $\ell$. Usually one assumes $\ell\sim 1$. 
For a long tube, $\ell\gg 1$, the \emph{spatially constant} modes,
$p_0\ne 0$, $\mathbf{p}=0$, become soft 
which means that one needs a larger box size $L$ to have a reasonable 
$\epsilon$--expansion. This is manifested by the fact that the
shape coefficients are increasing functions of the aspect ratio $\ell$.

In the $\delta$--expansion one separates the dynamics of the
spatially constant modes described by the O($n$) rotator, with 
the energies $\order{F^{-2}\Ls^{-3}}$.
For $\Lt \gg \Ls/(2\pi)$ the $\mathbf{p}\ne 0$ modes are exponentially
suppressed in the partition function, hence the regions of validity
of the $\epsilon$-- and $\delta$--expansions overlap in this case.
The $\mathbf{p}\ne 0$ states are responsible for exponentially 
small corrections $\sim \exp(-c \ell)$. 
However, a possible distortion of the higher lying rotator 
spectrum can still produce a  correction of type $\sim \ell^{-\nu}$
to the result obtained by using the ``standard'' rotator spectrum 
$\propto \mathcal{C}_{n;l}$. 
This is discussed in section~\ref{Distortion}.

The transition to the $\delta$-regime involves relations between
the 4d shape coefficients for $\ell\gg 1$ and the 3d shape coefficients
in a cubic box, as given e.g. in \eqref{beta1rel} and 
\eqref{rho_rel}-\eqref{gamma3_rel}.

In subsection \ref{PTmass} we compute the susceptibility
for small quark masses (i.e. small $M$) in the framework of $\chi$PT and 
subsequently use the comparison to the result \eqref{chi_rotm} 
to determine the coefficient of $M^2$ in $\eta$ 
as function of $F \Ls$ to NNLO.  

\subsection{The symmetric case}
\label{symmcase}

From the mass gap calculation in \cite{Niedermayer:2016yll} we obtained
the moment of inertia:
\begin{equation}
  \Theta= F^2\Ls^3 \left[ 1 + \frac{\widetilde\Theta_1}{F^2 \Ls^2}
    + \frac{\widetilde\Theta_2}{F^4 \Ls^4}+\ldots \right]\,,
  \label{Theta}
\end{equation}
with
\begin{align}
  \widetilde\Theta_1 &= (n-2)\beta_1^{(3)} \,, 
  \label{Tt1}
  \\
  \widetilde\Theta_2 
  &= (n-2)\left[\beta_1^{(3)}\left(\beta_1^{(3)} + \frac34\right)
    - 2 c_w\right]
  \nonumber
  \\
  & \quad
  - \frac{5(n-2)\rho}{12\pi^2} \log(\cov \Ls M_\pi)
  - 4\rho \left( 2\lr{1}+n\lr{2} \right)\,.
  \label{Tt2}
\end{align}
Here $\beta_1^{(3)}\equiv \beta_1^{(3)}(1)$ is a shape function
corresponding to a cubic 3d volume,  
$\rho= 8\pi^2 \beta_2^{(3)}(1)$ \eqref{rho_rel};
for notations undefined here we refer the reader
to \cite{Hasenfratz:1989pk} and \cite{Niedermayer:2016yll}.
Further $\lr{i}$ are renormalized low energy constants (LEC)
of Gasser and Leutwyler \cite{Gas84}; 
\begin{equation}
  \ln\overline{c}=-\frac12\left[\ln(4\pi)-\gamma_E+1\right]=-1.476904292\,,
  \label{ovC}
\end{equation}
and the constant $c_w$ was determined in \cite{Niedermayer:2016ilf} as
\begin{equation}
c_w=0.0986829798\,,
\label{cw}
\end{equation}
differing slightly from the original computation 
of Hasenfratz \cite{Hasenfratz:2009mp}.

Inserting \eqref{Theta} into \eqref{chi_rot0} one obtains
\begin{equation}
  \begin{split}
    \chi &= \frac{2}{n} F^2 \left[ 1 
      +\frac{1}{F^2 \Ls^2}\left(\widetilde\Theta_1-\frac{\ell}{6}(n-2)\right)
      \right.
      \\
      & \quad \left.  +\frac{1}{F^4\Ls^4}\left(\widetilde\Theta_2
        +\frac{\ell^2}{180}(n-2)(n-4) \right)
      +\ldots\right] \,. 
  \end{split}
  \label{chiTheta}
\end{equation}

On the other hand the susceptibility calculated by {$\chi$}PT
is given by \cite{Niedermayer:2016yll} 
\begin{equation}
  \chi = \frac{2}{n} F^2 \left[ 1 + \frac{\widetilde{R}_1}{F^2\Ls^2}
    + \frac{\widetilde{R}_2}{F^4\Ls^4}
    +\ldots\right]\,,
  \label{chi_PT}
\end{equation}
with
\begin{equation}
  \widetilde{R}_1 = -\frac{(n-2)}{4\pi}\left(\gamma_2-1\right)\,,
  \label{R1_On_resA}
\end{equation}
and
\begin{equation}
  \begin{split}
    \widetilde{R}_2 & = -\frac{(n-2)}{16\pi^2}
    \left[\left(\gamma_2-1\right)^2 +8\pi\left(\gamma_2-1\right)\beta_1
      + 2 \mcWov-\frac{(n-2)}{\ell}\gamma_3
    \right]
    \\
    & 
    + \frac{(n-2)}{24\pi^2}\left[\frac{3n-7}{\ell} 
      - 5 \left(\gamma_1 - \frac12\right) \right]
    \log\left(\cov L M_\pi\right) 
    \\
    & 
    -2  \left(2\lr{1}+n\lr{2}\right)\left(\gamma_1-\frac12\right)
    +4 \left[(n-1) \lr{1}+\lr{2}\right] \frac{1}{\ell}\,.
  \end{split}
  \label{R2_On_resA}
\end{equation}

The next-to-leading (NLO) and NNLO terms in \eqref{chiTheta} and \eqref{chi_PT} 
agree for large $\ell$ provided that  
\begin{align} \label{checkNLO}
  &\widetilde{R}_1 + \frac{\ell}{6}(n-2) \simeq \widetilde\Theta_1 \,,
\\  \label{checkNNLO}
  &\widetilde{R}_2 - \frac{\ell^2}{180} (n-2)(n-4) \simeq \widetilde\Theta_2 \,.
\end{align}
These involve relations between the 4d shape functions in the long cylinder
($\ell\gg 1$) to the shape functions for a 3d cube. 
Derivations of the relations required are supplied in App.~\ref{AppShC}. 

Firstly at NLO, using the large $\ell$ behavior of $\gamma_2$ given in
\eqref{gamma2_rel}, one indeed verifies \eqref{checkNLO}.

Next, the three relations \eqref{beta1rel}, \eqref{rho_rel},
\eqref{gamma3_rel} for the 1-loop shape functions
together with a fourth relation \eqref{R2T2rel}
referring to the 2-loop sunset diagram, ensure the validity of the
NNLO consistency condition \eqref{checkNNLO}.
This agreement provides, in our opinion, 
a highly non-trivial check on the various 
rather technical computations.

\subsection{Perturbative computation of the isospin 
  susceptibility for O($n$) with a mass term using 
  dimensional regularization}
\label{PTmass}

Here we restrict attention to a $\Ls^{d-1}\times\Lt\,,d=4$ volume
with periodic boundary conditions in all directions.
For perturbative computations we employ dimensional regularization
and add further $D-4$ dimensions of size $\Lhat=\Ls$. 
In ref.~\cite{Niedermayer:2016yll} we did not fix the value of
$\ellhat\equiv\Lhat/\Ls$; confirming independence of physical 
quantities on $\ellhat$ serves as an additional useful check 
of the computations.   

\subsubsection{The effective action with mass terms}

The effective action is given by
\begin{equation}
A = A_0 + A_M\,,
\end{equation}
where $A_M$ vanishes in the chiral limit $M=0$, with
\begin{align}
A_0&=\sum_{r=1}A_{2r}\,, 
\\
A_M&=\sum_{r=1}A_{M;2r}\,.
\end{align}

The lowest order actions are ($\bS(x)^2=1$) 
\begin{align} \label{A_cont0}
  A_2 &= \frac{1}{2 g_0^2} \int_x \sum_{\mu} 
    (\pmu \bS(x) \pmu \bS(x))\,,
\\
  A_{M;2}&=-\frac{M^2}{g_0^2}\int_x S_n(x)\,.
\end{align}

There are two independent $4$--derivative terms in the massless case: 
\begin{equation}
  A_4=\sum_{i=2,3}\frac{g_4^{(i)}}{4}A_4^{(i)}\,,
  \label{A4total_cont}
\end{equation}
with the $A_4^{(i)}$ given in App.~\ref{action_contribs}.

Terms involving the mass parameter in the next order are 
\begin{equation}
  A_{M;4}=\sum_{i=1,2} L_i A_{M;4}^{(i)}\,,
  \label{A4Mtotal_cont}
\end{equation}
with
\begin{align}
  A_{M;4}^{(1)}&=M^4 \int_x S_n(x)^2\,,
  \label{A4M1cont}
\\
  A_{M;4}^{(2)}&=\frac12 M^2 \int_x S_n(x)
 \sum_\mu (\pmu \bS(x) \pmu \bS(x))\,.
\end{align}
Actually one of the terms above is redundant for physical quantities.
To see this
consider the infinitesimal change of variables (preserving $\bS^2=1$): 
\begin{equation}
\bS \to \bS+\epsilon\delta(\bS)\,,
\end{equation}
with
\begin{align}
\delta(S_n)&=-\sum_{j=1}^{n-1}S_j^2\,,
\\
\delta(S_i)&= S_i S_n\,,\,\,\,\,i\ne n\,.
\end{align}
Under this change the leading $\chi$PT action changes according to
\begin{equation}
\delta\left(A_2+A_{M;2}\right) = 
\frac{1}{M^2g_0^2}\left[-A_{M;4}^{(1)}+2A_{M;4}^{(2)}\right]\,.
\end{equation}
Rearranging $A_{M;4}$
\begin{equation}
L_1A_{M;4}^{(1)}+L_2A_{M;4}^{(2)}=
\frac12\left[2L_1+L_2\right]A_{M;4}^{(1)}
+\frac12 L_2\left[-A_{M;4}^{(1)}+2A_{M;4}^{(2)}\right]\,,
\end{equation}
we see that physical quantities should just depend on the combination
$2L_1+L_2$; this will serve as a check on the computed 
contributions coming from the $A_{M;4}^{(i)}$.

\subsubsection{The chemical potential with mass terms}

The chemical potential $h$ is introduced by 
\begin{equation}
  \partial_0 \to \partial_0 -h Q\,,
\end{equation}
where $(QS)_1=iS_2$, $(QS)_2=-iS_1$, $(QS)_3=\ldots=0$.
This gives the $h$-dependent part of the action:
$$
A_h+A_{Mh}\,,
$$
where the massless part has contributions
\begin{equation}
A_h=A_{2h}+A_{4h}+\dots
\end{equation}
Here we only consider terms up to and including order $h^2$:
\begin{align}\label{A2hA4h}
  A_{2h}&=ihB_2+h^2 C_2+\dots\,,
\,\,\,\,\,
  A_{4h}=ihB_4+h^2 C_4+\dots\,,
\\
  B_4&=\sum_{i=2,3}\frac{g_4^{(i)}}{4}B_4^{(i)}\,,
\,\,\,\,\,
  C_4=\sum_{i=2,3}\frac{g_4^{(i)}}{4}C_4^{(i)}\,.
\nonumber
\end{align}
The expressions for $B_2\,,C_2\,,B_4^{(i)}\,,C_4^{(i)}$
are given in App.~\ref{action_contribs}.

To lowest order for the mass term  
\begin{equation}
  A_{Mh}=ihB_{M;4}+h^2 C_{M;4}+\dots\,,
\end{equation}
with
\begin{align}
  B_{M;4} & = -M^2 L_2\int_x\,S_n(x) j_0(x)\,,
\\
  C_{M;4} & = \frac{1}{2}M^2 L_2\int_x S_n(x)[Q\bS(x)]^2\,,
\end{align}
where $j_\mu(x)$ is defined in \eqref{jmu}.

The $h$--dependent part of the free energy $f_h$ is given by
($V_D=\Lt\Ls^{D-1}$):
\begin{align}
\mre^{- V_D f_h}&=\frac{Z(h)}{Z(0)} 
=\frac{\langle \mre^{-A_h-A_M-A_{Mh}} \rangle}{\langle\mre^{-A_M} \rangle}
\nonumber\\
&=1+\left[1+\langle A_M \rangle+\langle A_M\rangle^2
-\frac12\langle A_M^2 \rangle\right]
\left[-\langle A_h \rangle+\frac12\langle A_h^2\rangle\right]
\nonumber\\
&+\left[1+\langle A_M \rangle\right]
\left[\langle A_h A_M \rangle-\frac12 \langle A_h^2 A_M\rangle
-\langle A_{Mh}\rangle+\langle A_h A_{Mh} \rangle \right]
\nonumber\\
&-\frac12 \langle A_h A_M^2\rangle+\frac14\langle A_h^2 A_M^2\rangle
\nonumber\\
&+\langle A_M A_{Mh} \rangle-\langle A_h A_M A_{Mh} \rangle
+\frac12\langle A_{Mh}^2 \rangle+\mrO(M^6,h^3)\,.
\end{align}
where, $\langle ...\rangle$ denotes correlation functions with the 
massless action, and we have only kept terms up to order 
$M^4$ and $h^2$. So for the free energy we have
\begin{align}
- V_D f_h&=\left[1+\langle A_M \rangle+\langle A_M\rangle^2
-\frac12\langle A_M^2 \rangle\right]
\left[-\langle A_h \rangle+\frac12\langle A_h^2\rangle\right]
\nonumber\\
&+\left[1+\langle A_M \rangle\right]
\left[\langle A_h A_M \rangle-\frac12 \langle A_h^2 A_M\rangle
-\langle A_{Mh}\rangle+\langle A_h A_{Mh} \rangle \right]
\nonumber\\
&-\frac12 \langle A_h A_M^2\rangle+\frac14\langle A_h^2 A_M^2\rangle
+\langle A_M A_{Mh} \rangle-\langle A_h A_M A_{Mh} \rangle
+\frac12\langle A_{Mh}^2 \rangle
\nonumber\\
&+\langle A_h \rangle\left[
\langle A_h A_M \rangle-\langle A_{Mh}\rangle\right]
-\frac12\langle A_h\rangle^2\left[1+2\langle A_M \rangle
+3\langle A_M\rangle^2-\langle A_M^2 \rangle\right]
\nonumber\\
&+\langle A_h \rangle\left[
2\langle A_M \rangle\langle A_h A_M \rangle
-2\langle A_M \rangle\langle A_{Mh}\rangle
-\frac12 \langle A_h A_M^2\rangle
+\langle A_M A_{Mh} \rangle\right]
\nonumber\\
&-\frac12\left[\langle A_h A_M \rangle-\langle A_{Mh}\rangle\right]^2
+\mrO(M^6,h^3)\,.
\end{align}
Noting that
\begin{equation}
\langle A_h\rangle=\mrO(h^2)\,,
\end{equation}
and that correlation functions
with an odd number of spins $S$ vanish: 
\begin{equation}
\langle A_{M;2}\rangle=0=\langle A_{M;4}^{(2)}\rangle\,,
\end{equation}
many contributions drop out so that to the order we are considering we have
\begin{align} \label{fhx}
- V_D f_h&=\left[1+L_1\langle A_{M;4}^{(1)}\rangle
-\frac12\langle A_{M;2}^2 \rangle-L_2\langle A_{M;2}A_{M;4}^{(2)}\rangle\right]
\left[-\langle A_h \rangle+\frac12\langle A_h^2\rangle\right]
\nonumber\\
&-\frac{h^2}{2}\langle C_2 A_{M;2}^2\rangle
-\frac{h^2}{4}\langle B_2^2 A_{M;2}^2\rangle
\nonumber\\
&+h^2L_1\langle C_2 A_{M;4}^{(1)}\rangle
-\frac{h^2}{2}\langle C_4 A_{M;2}^2\rangle
-h^2L_2\langle C_2 A_{M;2}A_{M;4}^{(2)}\rangle
+h^2\langle A_{M;2}C_{M;4} \rangle
\nonumber\\
&-\frac{h^2}{2}L_1\langle B_2^2 A_{M;4}^{(1)}\rangle
-\frac{h^2}{2}L_2\langle B_2^2 A_{M;2}A_{M;4}^{(2)}\rangle
\nonumber\\
&-\frac{h^2}{2}\langle B_2B_4 A_{M;2}^2\rangle
+h^2\langle B_2 A_{M;2}B_{M;4} \rangle
+\mrO(M^6,h^3)+\mathrm{ho}\,,
\end{align}
where ``$\mathrm{ho}$" stands for higher order terms 
in the chiral expansion. 
In particular we drop terms $\mrO(M^4g_0^0)$.

The terms appearing in the averages above are not O($n$)
invariant, and so before starting the perturbative computations
we set $\bS(x) = \Omega \bR(x)$ and average over the rotations 
$\Omega$. The resulting expressions are given in App.~\ref{rot_averages}.

The free energy has a small $h$ expansion
\begin{equation} \label{fh}
f_h = -\frac12 h^2\chi(M)+\mrO\left( h^4 \right)\,.
\end{equation}
Using the results in App.~\ref{rot_averages}
the uniform susceptibility (at $h=0$) is given by
\begin{equation} \label{chi}
\chi(M) =\chi_0+\frac{M^4V_D^2}{g_0^4}\chi_1+\mrO\left(M^6\right)\,,
\end{equation}
where $\chi_0$ is the susceptibility for $M=0$ which we computed
previously \cite{Niedermayer:2016yll}:
\begin{equation}
\chi_0=\frac{2}{n g_0^2} -\frac{4 }{n(n-1) g_0^4}\langle W \rangle
-\frac{2}{V_D}\langle C_4 \rangle\,,
\end{equation}
where $W$ is defined in \eqref{W} and $\langle C_4^{(i)}\rangle$
in \eqref{C42av} and \eqref{C43av}.

Next, collecting the terms in App.~\ref{rot_averages}, 
\begin{align}
\chi_1&=\frac{1}{2n}
\left[-1-P_1+\frac{2}{V_D}g_0^4 L_1+g_0^2 L_2 P_4\right]\chi_0
\nonumber\\ \label{chi1}
&+\frac{1}{n(n-1)(n+2)g_0^2}
\left[n-1+(n-3)P_1-2P_2-\frac{1}{(n-2)}P_3\right]
\nonumber\\
&+\frac{1}{n(n+2)V_D}\left[-2 g_0^2 L_1
-n\langle C_4\rangle-L_2\int_z\langle (\pmu\bR(z)\pmu\bR(z))\rangle
+2g_0^2 L_2\right]+\mrO(g_0^4)\,,
\end{align}
where $P_1,P_2,P_3,P_4$ are given in \eqref{P1},\eqref{P2},\eqref{P3},
\eqref{P4} respectively.

Now we can begin with
the perturbative computations proceeding as usual by first
separating the zero mode and then changing to $\vpi$ variables
according to $\bR = (g_0 \vpi, \sqrt{1-g_0^2 \vpi^2})$.
Details of the perturbative computations to NNLO are
given in App.~\ref{pert_calc}.

Summing the terms (for $d=4$) one obtains for 
$\chi_0$ the result \eqref{chi_PT}, (note $F^2=1/g_0^2\,$).

For $\chi_1$ it is shown that it has a perturbative expansion of the form
\begin{equation}
\chi_1=-\frac{2}{n^2(n+2)}F^2\left[1+\frac{
\widetilde{\chi}_1^{(1)}}{F^2\Ls^2}
+\frac{\widetilde{\chi}_1^{(2)}}{F^4\Ls^4}+\ldots\right]\,.
\end{equation}
Here the NLO coefficient is  
\begin{equation}
\widetilde{\chi}_1^{(1)}=(2n-1)\beta_1+\frac{1}{2\pi}(\gamma_2-1)\,.
\label{chi11_pert}
\end{equation}
Using the behavior of the shape functions $\beta_1$ and $\gamma_2$
given in \eqref{beta1rel} and \eqref{rho_rel} we have for large $\ell$:
\begin{equation}
\widetilde{\chi}_1^{(1)}\simeq 
(2n-3)\beta_1^{(3)}(1)-\frac{1}{12}(2n-5)\ell\,.
\label{chi11pert_large_ell}
\end{equation}

At the NNLO we first show that in order
to cancel the $1/(D-4)$ pole terms we must have
\begin{equation}\label{L1L2}
  L_1+\frac12 L_2=\frac{(n-3)}{32\pi^2}\left[
    \frac{1}{D-4}+\ln\left(\overline{c}\Lambda_3\right)\right]\,.
\end{equation}
This agrees with the result of Gasser and Leutwyler \cite{Gas84} for $n=4$
if we set $L_1=-l_3\,,L_2=0$.
Then
\begin{align}
\widetilde{\chi}_1^{(2)}&=\widetilde{R}_2+\frac{1}{64\pi^2}
\left[4n(\gamma_2-1)^2
+32(2n-1)(n+1)\pi^2\beta_1^2+32(2n-1)\pi\beta_1(\gamma_2-1)\right. 
\nonumber\\
&+\frac{2}{\ell}\left\{(4n^2-5n+3)\alpha_2  
- 4n(n-1)\gamma_3- 2(n-3)\ln\left(\overline{c}\Lambda_3 \Ls\right)\right\}
\nonumber\\
&\left.+(5n-4n^2-3)\frac{1}{\ell^2}\right]\,.
\label{chi12_pert}
\end{align}

Using the expansions \eqref{alpha_ell} (for $s=2\,,d=4$) and
\eqref{gamma3_rel} we have for large $\ell$:
\begin{align}
\widetilde{\chi}_1^{(2)}&\simeq
\widetilde{\Theta}_2
+\frac12(2n-3)(n-1)\beta_1^{(3)}(1)^2
\nonumber\\
&-\frac{1}{12}(2n-5)(n-1)\beta_1^{(3)}(1)\ell
+\frac{1}{240}(n^2-12n+17)\ell^2
\nonumber\\
&+\frac{(n-3)}{16\pi^2\ell}\left\{\frac13-\ln(\overline{c}\Lambda_3\Ls)
-\frac12\alpha_0^{(3)}(1)\right\}\,.
\label{chi12pert_large_ell}
\end{align}

Agreement of the rotator result \eqref{chi_rotm}
with the perturbative result for large $\ell$ above requires 
that the parameter $\eta$ in the rotator Hamiltonian \eqref{HMh} 
has a $\chi$PT expansion of the form: 
\begin{equation}
\eta=M^2F^2\Ls^3\left[1+\frac{\mathcal{Z}_1}{F^2\Ls^2}
+\frac{\mathcal{Z}_2}{F^4\Ls^4}+\dots\right] +\order{M^3} \,,
\end{equation}
with 
\begin{align}
\mathcal{Z}_1&=\frac12(n-1)\beta_1^{(3)}(1)\,,
\\
\mathcal{Z}_2&=-\frac18(n-1)(n-3)\beta_1^{(3)}(1)^2\,.
\end{align}

\section{Distortion of the rotator spectrum}
\label{Distortion}

It is expected that at some higher order the standard rotator spectrum
$E_l \propto\mathcal{C}_{n;l}$ will be modified. It is interesting that
by comparing the NNLO results for the isospin susceptibility obtained
from $\chi$PT, eq.~\eqref{chi_PT}, and 
$\chi_{\text{rot}}$ given by \eqref{chi_rot0}, 
one can answer this question under reasonable assumptions.

The two results in NNLO differ by $\propto 1/\ell$ terms for $\ell\gg 1$,
\begin{equation} \label{dchi}
    \frac{\chi - \chi_{\text{rot}}}{\chi} =
    \frac{1}{F^4 L_s^4}\left( \frac{\Delta_2}{\ell} + \ldots\right)
    +\order{\frac{1}{F^{6} L_s^{6}}}
\end{equation}
where 
\begin{equation} \label{Delta2}
  \Delta_2  = 4(n+1) \left[ \lr{1} + \lr{2} + \frac{n-2}{32\pi^2}
    \left( 
      \log(\cov L_s M_{\pi}) + \frac12 \alpha_0^{(3)}(1) - \frac13
    \right) \right] \,.
\end{equation}
The omitted terms in \eqref{dchi} at the given order are 
vanishing exponentially fast for $\ell\to\infty$.\footnote{%
Note that the $\mathbf{p}\ne 0$ modes yield such contributions.}

The deviation \eqref{dchi} means that the spectrum of the spatially 
constant modes, the true rotator spectrum,
is not given exactly by the standard 
rotator spectrum $E_l=\mathcal{C}_{n;l}/(2\Theta)$, 
it is distorted already at energies $E_l \ll 1/L_s$. 

Assuming that the distortion has the form
\begin{equation} \label{deltaE}
  \delta E_l = \frac{\alpha}{L_s}(\mathcal{C}_{n;l})^\kappa \,,
\end{equation}
one can calculate the leading correction to the susceptibility
$\chi_{\text{rot}}$ of the standard rotator.

One has\footnote{%
Note that $w_1(l)=w_0(l)\mathcal{C}_{n;l}/d_n$
where $d_n=n(n-1)/2=\mathrm{dim}(\mathrm{O}(n))$.}
\begin{equation} 
  \begin{aligned} 
    z_1(u) & = \frac{2}{n(n-1)} 
    \left(-\frac{\partial }{\partial u}\right) z_0(u) \,,
    \\
    \delta z_j(u) & = - \alpha \ell 
    \left(-\frac{\partial}{\partial u}\right)^\kappa  z_j(u) \,,\quad  j=0,1 \,.
  \end{aligned} 
\end{equation}

From \eqref{z0} the leading $u\to 0$ behavior of the partition function is
$z_0(u)\propto u^{-a}$ with $a=(n-1)/2$.
This gives for the leading contribution
\begin{equation} 
  \begin{aligned} 
  \frac{\chi-\chi_{\text{rot}}}{\chi} 
  &= \frac{\delta z_1(u)}{z_1(u)} - \frac{\delta z_0(u)}{z_0(u)}
  = \kappa\alpha \ell u^{-\kappa} (a+1)\ldots(a+\kappa-1)  + \ldots
  \\
  &= 2^{\kappa} \kappa\alpha \frac{(F L_s)^{2\kappa}}{\ell^{\kappa-1}} 
  (a+1)(a+2)\ldots(a+\kappa-1) + \ldots
  \end{aligned} 
\end{equation}
(The omitted terms are higher order in $1/(F^2 L_s^2)$.)
The observed deviation \eqref{dchi} requires $\kappa=2$ and 
$\alpha\propto 1/(F L_s)^8$.
Finally, for the coefficient of the distortion one obtains
\begin{equation} \label{alpha}
  \alpha = \frac{1}{(FL_s)^8}\left[
    \lr{1} + \lr{2} 
    + \frac{n-2}{32\pi^2}\left( \log(\cov L_s M_{\pi})
      + \frac12 \alpha_0^{(3)}(1) - \frac13 \right) 
  \right] \,.
\end{equation}
The distortion becomes comparable with the leading term at
the isospin quantum number $l \sim (F L_s)^3 \gg 1$.

Note that the NNLO calculation for the mass gap means
determining $L_s m_1$ up to (and including) the $(F L_s)^{-6}$
terms. On the other side from eqs.~\eqref{deltaE}, \eqref{alpha} 
the obtained correction is $\sim (F L_s)^{-8}$, i.e. a NNNLO term.
The reason is that the typical values of the Casimir invariant 
in the partition function are $\mathcal{C}_{n;l} \sim F^2 L_s^2/\ell$.
As a consequence the distortion \eqref{deltaE} changes 
the susceptibility by $(F L_s)^{-4}/\ell$, i.e. in NNL order.

\begin{appendix}
  
  \section{\boldmath The effective O($n$) rotator}
  \label{AppRot}
  
  \subsection{Spectrum and multiplicities}
  \subsubsection{Symmetric case}
  \label{AppRot11}  

The energy levels of the Hamiltonian \eqref{H0h} are
\begin{equation}
  E^{(n)}_{lm} = \frac{1}{2\Theta}\mathcal{C}_{n;l} - h m \,.
  \label{Elm}
\end{equation}
Here $m$ is the quantum number associated with rotation in the 12-plane,
with values $m=-l,\ldots,l$.

The subspace of states corresponding to a given energy level
can be decomposed according to their transformation properties 
under the O($n-1$) transformations affecting the first $n-1$
components.
 
The multiplicity of the energy level \eqref{Elm}, $g_{lm}^{(n)}$ satisfies 
the recursion relation
\begin{equation}
  g^{(n)}_{lm} = \sum_{k=|m|}^{l} g^{(n-1)}_{km}\,.
  \label{glm_rr}
\end{equation}
For $n=3,4$ one has (assuming $|m|\le l$)
\begin{equation}
  \begin{aligned}
    g^{(3)}_{lm} &= 1 \,, \\
    g^{(4)}_{lm} &= l-|m|+1 \,.
  \end{aligned}
  \label{glm34}
\end{equation}
The solution of \eqref{glm_rr}, \eqref{glm34} for general $n$ 
is given by
\begin{equation}
  g^{(n)}_{lm} = \frac{(l-|m|+n-3)!}{(n-3)! (l-|m|)!}\,.
  \label{glm}
\end{equation}
The total multiplicity of states with a given $l$ is
\begin{equation}
  g_l^{(n)} = \sum_{m=-l}^{l} g^{(n)}_{lm} 
  = \frac{(l+n-3)!}{(n-2)! \; l!}(2l+n-2) \,.
  \label{gl}
\end{equation}

\subsubsection{Rotator in external field}
  \label{AppRot12}  
Consider now the O($n$) rotator in a small external magnetic field,
with Hamiltonian given in \eqref{HMh}. For simplicity we take here $h=0$
since its effect is simply adding $-hm$ to the energy levels,
as in \eqref{Elm}.

The external field splits the energy levels 
$E^{(n)}(l)=\mathcal{C}_{n;l}/(2\Theta)$
corresponding to a given O($n$) isospin $l$  into levels characterized
by the O($n-1$) isospin $k$, where $k=0,\ldots,l$.

To determine the expansion of the energy levels for small breaking
parameter $\eta$ one needs the transition matrix elements
between the corresponding eigenstates
\footnote{unit normalized $\langle l,k,\alpha | l,k,\alpha \rangle = 1$}.
These are
\begin{equation}
 v_{lk}^{(n)} = \langle l+1,k,\alpha | \Snhat| l,k,\alpha\rangle =
 \sqrt{\frac{(l+1-k)(l+n-2+k)}{(2l+n-2)(2l+n)}}\,, \quad 0\le k \le l \,.
 \label{v_lk}
\end{equation}
Here $\alpha$ denotes the remaining quantum numbers besides $k$ 
characterizing an O($n-1$) eigenvector. (For $n=3$ one has $k=|m|$
and $\alpha$ is the sign of $m$.)
The leading order PT gives
\begin{equation}
  E^{(n)}(l,k) = E^{(n)}(l) + \epsilon^{(n)}(l,k)\eta^2 \Theta 
  + \order{\eta^4\Theta^3}\,,
 \label{E_lk}
\end{equation}
where
\begin{equation}
  \epsilon^{(n)}(l,k) = \rho^{(n)}(l-1,k) - \rho^{(n)}(l,k) \,, 
  \label{eps_lk}
\end{equation}
and
\begin{equation}
  \rho^{(n)}(l,k) = \frac{2(l+1-k)(l+n-2+k)}{(2l+n-2)(2l+n-1)(2l+n)}\,.
  \label{rho_lk}
\end{equation}

Accordingly the energy of the ground state becomes
 \begin{equation}
  E^{(n)}(0,0) = -\frac{2}{n(n-1)} \eta^2 \Theta\,,
  \label{En_00}
\end{equation}
while the $n$--plet of O($n$) is split into a singlet and $(n-1)$--plet
under O($n-1$):
\begin{equation}
  \begin{aligned}
    E^{(n)}(1,0) &=
    \frac{n-1}{2\Theta}-\frac{2(n-7)}{(n+2)(n^2-1)}\eta^2\Theta\,,
    \\
    E^{(n)}(1,1) &= \frac{n-1}{2\Theta} -\frac{2}{(n+1)(n+2)} \eta^2\Theta\,.
  \end{aligned}
  \label{En_1011}
\end{equation}
The lower one is the $(n-1)$--plet.
For $n=4$ one obtains for the mass gap 
$E(1,1)-E(0,0)=3/(2\Theta)[1+\eta^2\Theta^2/15+\ldots]$, 
in agreement with \cite{Leutwyler:1987ak,Weingart:2010yv}.

A useful check for the transition matrix element
and the multiplicity is to calculate the trace of $\Snhat^2$
in the O($n$) invariant subspace with given $l$ which should 
give $g^{(n)}_l/n$. Written out explicitly one has indeed
\begin{equation}
  \begin{aligned}
    \Tr(\Snhat^2 P_l) &= \sum_{k=0}^l \sum_\alpha 
    \langle l,k,\alpha|\Snhat^2|l,k,\alpha\rangle\\
    &= \sum_{k=0}^l g^{(n-1)}_k \left[ \left(v^{(n)}(l,k)\right)^2
      + \left(v^{(n)}(l-1,k)\right)^2\right]
    = \frac{1}{n} g^{(n)}_l
    \,.
  \end{aligned}
  \label{TrS0sq}
\end{equation}

\subsection{The transition matrix elements}

The squared angular momentum operator in $n$ dimensions, $\hat{L}^2$
is given (up to the sign) by the angular Laplacian $\Delta_{S^{n-1}}$
on the unit sphere $S^{n-1}$. The eigenfunctions of $\hat{L}^2$
corresponding to the eigenvalue $\mathcal{C}_{n;l}$ can be written as
\begin{equation}
  \Psi_l(\phi_1,\ldots,\phi_{n-2},\theta) 
  = f_{lk}(\theta) g_k(\phi_1,\ldots,\phi_{n-2})
\end{equation}
where $g_k(\phi_1,\ldots,\phi_{n-2})$ is an eigenfunction of the
$n-1$ dimensional squared angular momentum operator 
$\sum_{i=1}^{n-1} \hat{L}_i^2$ with eigenvalue $k(k+n-3)$
and $f_{lk}(\theta)$ satisfies the differential equation
\begin{equation} \label{eqy}
  (1-x^2) y'' - (n-1) x y' + 
    \left[\mathcal{C}_{n;l} - \frac{k(k+n-3)}{1-x^2} \right] y = 0 \,,
\end{equation}
where $x=\cos\theta$ and $f_{lk}(\theta)=y(\cos\theta)\,.$

For $k=0$ the solution is given by the O($n$) Legendre polynomials
\begin{equation} \label{P_nl0}
  P_{nl0} = \frac{(-1)^l \Gamma\left( \frac{n-1}{2} \right)}%
    { 2^l \Gamma\left( l + \frac{n-1}{2} \right) } 
  (1-x^2)^{-(n-3)/2}\left( \frac{\mrd}{\mrd x} \right)^l
  (1-x^2)^{l+(n-3)/2} \,.
\end{equation}
The coefficient is chosen to satisfy the normalization condition 
$P_{nl0}(1)=1$.

For general $k$ the solution of the differential equation \eqref{eqy}
is given by the associated O($n$) Legendre functions:
\begin{equation} \label{P_nlk}
  P_{nlk} = (1-x^2)^{k/2}\left(\frac{\mrd}{\mrd x}\right)^k P_{nl0}(x) \,.
\end{equation}
Two of these functions corresponding to different values of $l$ 
(and same $n,k$) are orthogonal with the weight
\begin{equation} \label{weight}
  \int_{-1}^1 \mrd x\,(1-x^2)^{(n-3)/2} \ldots
\end{equation}
For convenience we introduce the notation
\begin{equation} \label{intw}
  \langle f(x) \rangle = 
  \int_{-1}^1 \mrd x\,(1-x^2)^{(n-3)/2} f(x) \,.
\end{equation}

The transition matrix element \eqref{v_lk} is then given by
\begin{equation}
  \frac{ \langle x P_{nlk}(x) P_{n,l+1,k}(x)\rangle}
  {\sqrt{\langle P_{nlk}(x)^2 \rangle \langle P_{n,l+1,k}(x)^2 \rangle}}
  = \sqrt{ \frac{(l+1-k)(l+n-2+k)}{(2l+n-2)(2l+n)} } \,.
\end{equation}

 \subsection{\boldmath Partition function of the O($n$) rotator}

\subsubsection{Symmetric case}
  \label{AppRot21}  

Here we consider the partition function \eqref{Z0h} of the O($n$)
rotator without an external magnetic field. 
Expanding it in the chemical potential, the expansion coefficients
are given by
\begin{equation}
  \begin{aligned}
    z_0(u) & = \sum_{l=0}^\infty w_0(l) \mre^{-u\mathcal{C}_{n;l}}
    = \frac{\Gamma\left((n-1)/2\right)}{\Gamma\left(n-1\right)}u^{-(n-1)/2}
    \\
    & \times 
    \left( 1 + \frac{u}{6}(n-1)(n-2)
      + \frac{u^2}{360}(n-1)(n-2)(5n^2-17n+18) +\ldots \right)\,, 
  \end{aligned}
  \label{z0}
\end{equation}
and
\begin{equation}
  \begin{aligned}
    z_1(u) & = \sum_{l=0}^\infty w_1(l) \mre^{-u\mathcal{C}_{n;l}}
    = 2\frac{\Gamma\left((n+1)/2\right)}{\Gamma\left(n+1\right)}u^{-(n+1)/2}
    \\
    & \times 
    \left( 1 + \frac{u}{6}(n-2)(n-3)
      + \frac{u^2}{360}(n-2)(n-5)(5n^2-17n+18) +\ldots \right)\,.
  \end{aligned}
  \label{z1}
\end{equation}
Here the weights $w_i(l)$ are (cf.~\eqref{gl})
\begin{align}
  w_0(l) &\equiv g_l^{(n)} \,,
  %= \sum_{m=-l}^{l} g^{(n)}_{lm} 
  %= \frac{(l+n-3)!}{(n-2)! \; l!}(2l+n-2) \,, 
  \label{w0}
  \\
  w_1(l) &= \sum_{m=-l}^{l} g^{(n)}_{lm} m^2
  = 2 \frac{(l+n-2)!}{n! (l-1)!} (2l+n-2) \,.
  \label{w1}
\end{align}

The coefficients appearing in \eqref{z0}, \eqref{z1} can be determined
numerically. The sum converges fast enough to separate the different
powers of $u$ and then obtain the corresponding polynomials in $n$,
``beyond a reasonable doubt''.
An exact calculation of the expansions is given in Appendix~\ref{AppZexp}.

\subsubsection{Rotator in external field}
  \label{AppRot22}

For calculating $z_i(u)$ to order $\eta^2$ and $h^2$ one also needs 
the weights
\begin{equation}
  \begin{aligned}
    w_2(l) &= \sum_{k=0}^l \epsilon^{(n)}(l,k) g^{(n-1)}_k \\
    &= - \frac{2(n-3)(2l+n-2)}{n(2l+n-3)(2l+n-1)}
   \frac{(l+n-3)!}{l!\, (n-2)!} \,,
  \end{aligned}
\end{equation}
and
\begin{equation}
  \begin{aligned}
    w_3(l) &= 
    \sum_{k=0}^l \sum_{m=-k}^k \epsilon^{(n)}(l,k)  g^{(n-1)}_{km} m^2 \\
    &= - \frac{4(n-1)(2l+n-2)}{(n+2)(2l+n-3)(2l+n-1)}
  \frac{(l+n-2)!}{(l-1)!\, n!} \,.
  \end{aligned}
\end{equation}

With these one obtains
\begin{equation}
  \begin{aligned}
    z_2(u) & = \sum_{l=0}^\infty w_2(l) \mre^{-u\mathcal{C}_{n;l}}
    = -2\frac{\Gamma\left((n+1)/2\right)}{\Gamma\left(n+1\right)}u^{-(n-3)/2}
    \\
    & \times 
    \left( 1 + \frac{u}{6}(n-1)(n-3)
      + \frac{u^2}{360}(n-1)(n-3)(5n^2-22n+18) +\ldots \right) \,, 
  \end{aligned}
  \label{z2}
\end{equation}
and
\begin{equation}
  \begin{aligned}
    z_3(u) & = \sum_{l=0}^\infty w_3(l) \mre^{-u\mathcal{C}_{n;l}}
    = -4\frac{\Gamma\left((n+3)/2\right)}{\Gamma\left(n+3\right)}u^{-(n-1)/2}
    \\
    & \times 
    \left( 1 + \frac{u}{6}(n-1)(n-5)
      + \frac{u^2}{360}(n-1)(n-7)(5n^2-22n+18) +\ldots \right) \,. 
  \end{aligned}
  \label{z3}
\end{equation}

For the partition function these give
\begin{equation}
  Z(h;\Theta,\eta) = z_0(u)\left[1  - \eta^2 \Theta \Lt r_2(u) 
    + \frac12 h^2 \Lt^2 \left(r_1(u) - \eta^2\Theta \Lt r_3(u)\right) 
    + \ldots \right]\,,
  \label{zua}
\end{equation}
where $r_i(u)=z_i(u)/z_0(u)$.

Finally, one obtains for the corresponding susceptibility 
\begin{equation}
    \chi = \frac{\Lt}{V_s}\Big\{
    r_1(u) - \eta^2\Theta \Lt\left[r_3(u)-r_1(u) r_2(u)\right]+\ldots\Big\}\,,
\end{equation}
from which \eqref{chi_rotm} follows.

\subsection{\boldmath Expansion of $z_i(u)$ for small $u$}
\label{AppZexp}

Consider the sum
\begin{equation}
f_\nu(u)=\sum_{k=1}^{\infty} k^\nu \mre^{-u k^2}\,.
\end{equation}
A useful representation to obtain the behavior as $u\to0$ is
\begin{equation} 
f_\nu(u)= \frac{1}{2\pi i}\int_{\sigma-i\infty}^{\sigma+i\infty}\mrd t\,
      u^{-t}\Gamma(t)\zeta(2t-\nu)
\end{equation}     
where $\sigma>(\nu+1)/2$ and $\zeta(s)$ is the Riemann zeta-function
\begin{equation}
\zeta(s)=\sum_{k=1}^{\infty}k^{-s}\,.
\end{equation}
$\zeta(s)$ has a pole at $s=1$ with residue 1. Also we will need
\begin{align}
\zeta(0)&=-\frac12\,,
\label{zeta0}
\\
\zeta(-n)&=-\frac{B_{n+1}}{n+1}\,,\,\,\,\, n\ge1\,,
\label{zeta-n}
\end{align}
where $B_n$ are the Bernoulli numbers
\begin{equation}
B_2=1/6\,,\,B_3=0\,,\,B_4=-1/30\,,\,B_5=0\,,\, B_6=1/42\,,\dots
\end{equation}
We will use the convention $B_1=1/2$ to have $\zeta(0)=-B_1$.

Also we note the Gamma function $\Gamma(t)$ has poles at  
$t=-n\,,\quad n=0,1,2,...$ with residue
\begin{equation}
  \mathop{\Res}_{t=-n} \Gamma(t) = \frac{(-1)^n}{n!}\,.
\end{equation}

Shifting the integration contour to the left we pick up the
residues of the poles, the first one at $t=(\nu+1)/2$ from
the $\zeta$ function and then poles at $t=0,-1,...$ from the
Gamma function. In this way we get 
(with the convention $B_1=1/2$) for small $u$
\begin{equation}
  f_\nu(u)\sim \frac12 \Gamma\left( \frac{\nu+1}{2} \right) u^{-(\nu+1)/2} 
  -\sum_{k=0}^\infty \frac{(-1)^k u^k}{k!(2k+\nu+1)} B_{2k+\nu+1}\,.
  \label{fnu_exp}
\end{equation}
Explicitly
\begin{equation}
  f_\nu(u)\sim \frac12 \Gamma\left( \frac{\nu+1}{2} \right) u^{-(\nu+1)/2}  
  - \frac{1}{\nu+1}B_{\nu+1}
  + \frac{u}{\nu+3}B_{\nu+3} - \frac{u^2}{2!(\nu+5)}B_{\nu+5} +\ldots
\end{equation}

For even $\nu$ the function $f_\nu(u)$ is related to the Jacobi 
theta-function and its derivatives. In particular
\begin{equation}
  S(u) = 1 + 2 f_0(\pi u) = u^{-1/2}\left(1+2\mre^{-\pi/u}+\ldots\right)\,.
\end{equation}
In this case there are no power corrections from the sum
in \eqref{fnu_exp}, except the $-1/2$ term for $k=\nu=0$.

\subsubsection{Even $n$ values}

Consider \eqref{w0} first for general integer $n$
introducing the variable $q=l+n/2-1$ and
\begin{equation} \label{tw0}
  \tilde{w}_0(q)= w_0(l) = 
  \frac{2q}{(n-2)!} (q-n/2+2)(q-n/2+3)\ldots (q+n/2-2) \,.
\end{equation}
It is even/odd according to the parity of $n$,
$\tilde{w}_0(-q)=(-1)^n \tilde{w}_0(q)$.
Expanding in powers of $q$ one has
\begin{equation}
\tilde{w}_0(q) = \sum_{r=1}^{\fl{n/2-1}}
  \tilde{\gamma}_{0,r} q^{n-2r}\,,
  \label{tgq}
\end{equation}
where $\fl{x}=\floor(x)$,
with the leading coefficients
\begin{equation}
  \tilde{\gamma}_{0,1} = \frac{2}{(n-2)!} \,, \quad
  \tilde{\gamma}_{0,2} = -\frac{1}{12 (n-5)!} \,, \quad
  \tilde{\gamma}_{0,3} = \frac{(5n-8)}{2880 (n-7)!} \,.
  \label{gammak}
\end{equation}

For the case when $n$ is even one has
\begin{equation} \label{z0u_even}
  \begin{aligned}
  z_0(u) & = \sum_{l=0}^\infty w_0(l) \mre^{-u\mathcal{C}_{n;l}}
  = \mre^{u\left(n/2-1\right)^2} \sum_{k=n/2-1}^\infty w_0(k-n/2+1) 
    \mre^{-u k^2}
  \\
  & = \mre^{u\left(n/2-1\right)^2} \sum_{k=1}^\infty \tilde{w}_0(k) 
    \mre^{-u k^2}
  = \mre^{u\left(n/2-1\right)^2} \sum_{r=1}^{n/2-1} 
    \tilde{\gamma}_{0,r} f_{n-2r}(u)\,.
  \end{aligned}
\end{equation}
Here we extended the summation range 
from  $k=n/2-1,\ldots,\infty$ to $k=1,\ldots,\infty$ 
observing that $\tilde{w}_0(k)=0$ for $k=1,2,\ldots,n/2-2$.
Finally, inserting \eqref{fnu_exp} and \eqref{gammak} we obtain 
the expansion \eqref{z0}.
The calculation goes the same way for $z_i(u)$, $i=1,2,3$  yielding 
the results stated in \eqref{z1}, \eqref{z2} and \eqref{z3}.

It is interesting to note that the properties listed above for 
$\tilde{w}_0(q)$ are essential for having a proper power series for $z_i(u)$,
otherwise one would obtain in the small--$u$ expansion a mixture of integer 
and half-integer powers.

\subsubsection{Odd $n$ values}

Define
\begin{equation}
  \begin{aligned}
    \bar{f}_\nu(u) & =\sum_{k=0}^{\infty} 
    \left( k +\frac12 \right)^\nu \mre^{-u(k+1/2)^2}
    \\
    &= 2^{-\nu}f_\nu(u/4) - f_\nu(u)\,. 
  \end{aligned}
\end{equation}
One has
\begin{equation}
  \begin{aligned}
  \bar{f}_\nu(u) & \sim 
  \frac12 \Gamma\left( \frac{\nu+1}{2} \right) u^{-(\nu+1)/2}
  + (1-2^{-\nu}) \frac{1}{\nu+1}B_{\nu+1}
  \\
  & \quad
  -(1-2^{-\nu-2}) \frac{u}{\nu+3} B_{\nu+3} 
  + (1-2^{-\nu-4})  \frac{2 u^2}{2!(\nu+5)}B_{\nu+5} +\ldots
  \end{aligned}
\end{equation}

In this case 
\begin{equation} \label{z0u_odd}
  \begin{aligned}
    z_0(u) &= \sum_{l=0}^\infty w_0(l) \mre^{-u\mathcal{C}_{n;l}}
    = \mre^{u\left(n/2-1\right)^2} \sum_{l=0}^\infty w_0(l) \mre^{-u(l+(n-3)/2 + 1/2)^2}
    \\
    & =  \mre^{u\left(n/2-1\right)^2} 
    \sum_{k=(n-3)/2}^\infty w_0(k+(n-3)/2+1/2) \mre^{-u(k + 1/2)^2}
    \\
    & = \mre^{u\left(n/2-1\right)^2} 
    \sum_{k=0}^\infty \tilde{w}_0(k+1/2) \mre^{-u(k + 1/2)^2}
    =  \mre^{u\left(n/2-1\right)^2} 
    \sum_{r=1}^{(n-1)/2} \tilde{\gamma}_{0,r} \bar{f}_{n-2r}(u)\,.
  \end{aligned}
\end{equation}
This yields the same analytic form \eqref{z0} as obtained
for even $n$ values.

\section{Shape coefficients for long 4d tube}
\label{AppShC}

In this appendix we calculate the behavior of the 
$1$-- and $2$--loop shape functions (appearing in our perturbative computations)
in $d=4$ dimensions for 
a long ``tube", $\Ls^3\times\Lt$ with $\ell\equiv \Lt/\Ls \gg 1\,$. 
They are related to the shape functions defined for 
a $3$--dimensional cube. 

We first recall some useful
relations involving the Jacobi theta--function and some properties of
the free massless propagator in an asymmetric periodic volume.

\subsection{Some properties Jacobi theta-function}

The Jacobi theta-function is defined by
\begin{equation}
  \begin{split}
    S(u,z) & = \sum_{n=-\infty}^\infty \mre^{-\pi u (n+z)^2}\\
    & = u^{-1/2 }\sum_{n=-\infty}^\infty \mre^{-\pi n^2/u}
    \cos(2\pi n z) \,.
  \end{split}
  \label{Suz}
\end{equation}
The first sum above converges quickly for $u \ge 1$ while the 
second for $0 < u \le 1$. 
For small and large $u$ it is given by 
\begin{equation}
    S(u,z) = 
  \begin{cases} 
    u^{-1/2}  + \mrO(\mre^{-\pi / u})\,, & \text{for } u\to 0\,.
    \\
    \mre^{-\pi z^2 u} + \mrO(\mre^{-\pi u/4}) \,, & \text{for }
    u\to\infty\,, |z|\le 1/2 \,.
  \end{cases}
  \label{Suz_as}
\end{equation}

Defining 
\begin{equation}
S(u)\equiv S(u,0)\,,
\end{equation}
one has
\begin{equation}
  \int_{-1/2}^{1/2} \mrd z\, 
  S\left(\frac{1}{u},z\right)
  =  \sqrt{u} \,,
  \label{intS}
\end{equation}
\begin{equation}
  \int_{-1/2}^{1/2} \mrd z\, 
  S\left(\frac{1}{u},z\right) S\left(\frac{1}{v},z\right)
  =  \sqrt{uv} \, S(u+v) \,,
  \label{intSS}
\end{equation}
and
\begin{equation}
  \int_{-1/2}^{1/2} \mrd z\, 
  \ddot{S}\left(\frac{1}{u},z\right) S\left(\frac{1}{v},z\right)
  = 4\pi \sqrt{uv} \, S'(u+v) \,,
  \label{intddSS}
\end{equation}
where $\ddot{S}(u,z) = \partial_z^2 S(u,z)$ 
and $S'(u,z)=\partial_u S(u,z)$.

Some relations which are also used in the following:
\begin{equation}
  \begin{split}
  \int_0^\infty \mrd t\, t^{a-1} \left(S^{d}(t)-1\right)
  & = \int_0^\infty \mrd t\, t^{a-1} \left[S^{d}(t)\right]_{\text{sub}}
  -\frac{2}{d-2a}-\frac{1}{a} \\
  & = \alpha^{(d)}_a(1) -\frac{d}{a(d-2a)}\,,
  \end{split}
\end{equation}
since
\begin{equation}
  \int_0^1 \mrd t\, t^{a-1} \left( t^{-d/2} - 1\right) 
  = -\frac{2}{d-2a}-\frac{1}{a}\,.
\end{equation}
In particular
\begin{equation}
    \int_0^\infty \mrd t\, t^{a-1} \left(S^{3}(t)-1\right)
    = \alpha^{(3)}_a(1) -\frac{3}{a(3-2a)} 
    = \alpha^{(3)}_{3/2-a}(1) - \frac{3}{a(3-2a)} \,.
\end{equation}
For  $D\sim 4$, $a=3/2$ one obtains a pole term 
\begin{equation}
    \int_0^\infty \mrd t\, t^{1/2} \left(S^{D-1}(t)-1\right)
    = - \frac{2}{D-4} + \alpha^{(3)}_0(1) -\frac23 +\mrO(D-4)\,.
\end{equation}

\subsection{Some properties of the free massless propagator}

In this paper we employ dimensional regularization 
and add $q=D-4$ extra dimensions of length $\Lhat=\Ls\,.$
We introduce $L_0=\Lt$ and $L_\mu=\Ls\,,\,\mu\ge1$ and
the volume $V_D=\VdD\Lt$\,, $\VdD=\Ls^{D-1}\,.$

The massless propagator with periodic boundary conditions
in all directions is given by
\begin{equation}
  G(x)=\frac{1}{V_D}\psump\frac{\mre^{ipx}}{p^2}\,,
  \label{Gmass0}
\end{equation}
where the sum is over momenta
$p=2\pi(n_0/L_0,\dots,n_{D-1}/L_{D-1})\,,\,\,n_k\in\Z\,,$ 
and the prime on the sum means that the zero momentum is omitted:
$\sum'_p=\sum_{p\ne0}\,$.

For $G(x)$ and $\ddot{G}(x)$ we have the representations
\cite{Niedermayer:2016ilf} (3.55)
\begin{equation}
  \Ls^{D-2} G(x) = \frac{1}{4\pi}\int_0^\infty \mrd u \left. \left\{
      u^{-D/2} S\left(\frac{\ell^2}{u},z_0\right)
      \prod_{\mu=1}^{D-1} S\left(\frac{1}{u},z_\mu\right)
      - \frac{1}{\mcVD} \right\}\right|_{z_\nu=x_\nu/L_\nu}\,,
  \label{GxS}
\end{equation}
and
\begin{equation}
    \Ls^D \ddot{G}(x) = \frac{1}{4\pi \ell^2}\int_0^\infty \mrd u \left.
    u^{-D/2} \ddot{S}\left(\frac{\ell^2}{u},z_0\right)
    \prod_{\mu=1}^{D-1} S\left(\frac{1}{u},z_\mu\right)
    \right|_{z_\nu=x_\nu/L_\nu} \,.
  \label{ddGxS}
\end{equation}

To study the large $\ell$ behavior it is convenient to separate
the 1d propagator with periodic b.c. from $G(x)$:
\begin{equation}
  G(x) = G_1(x) + \mcR(x)\,,
\end{equation}
where
\begin{equation}
  G_1(x) = \frac{\Lt}{\VdD}\Delta_1\left(\frac{x_0}{\Lt}\right)\,,
  \qquad (|x_0|\le \Lt/2)\,,
\end{equation}
with the 1d propagator with pbc on the interval $z\in [-1/2,1/2]$,
\begin{equation}
  \Delta_1(z) 
  = -\frac12 |z| + \frac12 z^2 + \frac{1}{12} 
  =   \frac{1}{4\pi} \int_0^\infty \mrd u \, 
  \left[ u^{-1/2} S\left(\frac{1}{u},z\right) - 1\right] \,.
  \label{Delta1z}
\end{equation}
Next
\begin{equation}
  \mcR(x) = \sum_{m=-\infty}^{\infty} R(x_0+m\Lt,\bfx)\,,
\end{equation}
with (see eq.~(5.8) in \cite{Niedermayer:2010mx})
\begin{equation}
  R(x) = \frac{1}{2\VdD} \sum_{\bfp\ne 0}
  \frac{1}{\omega_{\bfp}}
  \mre^{-\omega_\bfp|x_0| } \mre^{i\bfp\bfx}\,,
  \qquad (\omega_\bfp=|\bfp|)\,.
  \label{Rxdef}
\end{equation}
The function $R(x)$ is defined in \eqref{Rxdef} for 
all $x\in\R^D\backslash0\,$; 
in particular for $|x_0|\to\infty$ the function $R(x)$ 
falls exponentially. The singularity at $x=0$ is 
regularized dimensionally through an alternative representation
in terms of the Jacobi theta function $S(u,z)$:
\begin{equation}
  \Ls^{D-2} R(x)=\frac{1}{4\pi}\int_0^\infty
  \frac{\mrd u}{\sqrt{u}} \mre^{-x_0^2\pi/(\Ls^2u)}  
  \left\{ u^{-(D-1)/2}
    \prod_{\mu=1}^{D-1}S\left(\frac{1}{u},\frac{x_\mu}{\Ls}\right)
    -1\right\}\,.
  \label{RxDR}
\end{equation}

\subsection{1-loop shape functions}

We start with the $1$--loop functions\footnote{Note $\beta_s$ is only defined
  for integer values of $s$.}
$\alpha_s^{(d)}(\ell)\,,\beta_s(\ell)\,,\gamma_s(\ell)$;
for notations undefined here we again refer the reader
to \cite{Hasenfratz:1989pk} and \cite{Niedermayer:2016yll}.

Neglecting terms decreasing exponentially fast with $\ell$ one has 
\begin{equation}
  \begin{split}
    & \alpha_s^{(d)}(\ell)  = \frac{1}{\ell}\int_0^\infty \mrd t\,
    t^{s-1}\left[ S\left( \frac{t}{\ell^2}\right) S^{d-1}(t) \right]_{\text{sub}}
    \\
    & \quad
    \equiv \frac{1}{\ell}\int_0^1 \mrd t\,
    t^{s-1}\left[ S\left( \frac{t}{\ell^2}\right) S^{d-1}(t) 
      - \ell\, t^{-d/2}\right]
    +  \frac{1}{\ell}\int_1^\infty \mrd t\,
    t^{s-1}\left[ S\left( \frac{t}{\ell^2}\right) S^{d-1}(t) - 1\right]
    \\
    & \quad
    \simeq \int_0^\infty \mrd t\, t^{s-3/2}\left[ S^{d-1}(t)\right]_{\text{sub}}
    + \frac{1}{\ell}\int_1^\infty \mrd t\,
    t^{s-1}\left[ S\left(\frac{t}{\ell^2}\right) -1\right]
    \\
    & \quad
    = \alpha_{s-1/2}^{(d-1)}(1) + 2\pi^{-s}\Gamma(s)\zeta(2s) \ell^{2s-1}
    -\frac{2}{2s-1} + \frac{1}{s \ell} \,.
  \end{split}
  \label{alpha_ell}
\end{equation}
The pole contributions at $s=0$ and $s=1/2$ cancel 
and one obtains in the corresponding limits
\begin{equation}
  \alpha_{1/2}^{(d)}(\ell) \simeq \alpha_{0}^{(d-1)}(1)
  + 2\ln(\ell) + \gamma_E - \ln(4\pi)+ \frac{2}{\ell}\,,
\end{equation}
\begin{equation}
  \alpha_{0}^{(d)}(\ell) \simeq \alpha_{-1/2}^{(d-1)}(1)
  + 2 - 2\frac{\ln(\ell)}{\ell} + 
  \left[\gamma_E - \ln(4\pi)\right] \frac{1}{\ell}\,.
\end{equation}
Note that $\alpha_{s-1/2}^{(d-1)}(1)=\alpha_{d/2-s}^{(d-1)}(1)$.

Now
\begin{equation}
  \beta_k^{(d)}(\ell) = \left(-\frac{1}{4\pi}\right)^k
  \left[ \alpha_k^{(d)}(\ell) + \frac{2}{2k-d} - \frac{1}{k\ell}
  \right]\,,\qquad(k\ne 0,d/2)\,.
  \label{beta_k}
\end{equation}
From this and \eqref{alpha_ell} we get for $d=4$, $k=1$: 
\begin{equation}
  \beta_1^{(4)}(\ell) + \frac{\ell}{12} \simeq \beta_1^{(3)}(1) \,.
  \label{beta1rel}
\end{equation}
This relation holds numerically extremely well already at $\ell=4$
where it gives $-0.10754837390 \simeq -0.10754837389$, 
illustrating the fact that the approach to the $\ell=\infty$ limit
is exponentially fast. The same comment will apply to 
similar relations obtained in the following.

For $\gamma_s(\ell)$ we obtain
\begin{equation}
  \begin{split}
    & \gamma_s^{(d)}(\ell)=-\frac{2}{\ell^3} \int_0^\infty \mrd t\, t^{s-1}
    \left[ S^{d-1}(t)S'\left( \frac{t}{\ell^2}\right)\right]_{\text{sub}}
    \\
    & \quad
    \equiv -\frac{2}{\ell^3}\int_0^1 \mrd t\,
    t^{s-1}\left[ S^{d-1}(t)S'\left( \frac{t}{\ell^2}\right)
      +\frac12 \ell^3\, t^{-d/2-1}\right]
    \\
    & \qquad\qquad
    -  \frac{2}{\ell^3}\int_1^\infty \mrd t\,
    t^{s-1} S^{d-1}(t)S'\left( \frac{t}{\ell^2}\right)
    \\
    & \quad
    \simeq \int_0^\infty \mrd t\, t^{s-3/2} \left[ S^{d-1}(t)\right]_{\text{sub}}
    -2 \ell^{2s-3} \int_{1/\ell^2}^\infty \mrd t\, t^{s-1} S'(t)
    \\
    & \quad
    = \alpha_{s-3/2}^{(d-1)}(1) 
    -2 \ell^{2s-3} \int_{1/\ell^2}^\infty \mrd t\, t^{s-1} S'(t) \,.
  \end{split}
  \label{gammas}
\end{equation}
Note that $\alpha_{s-3/2}^{(d-1)}(1)=\alpha_{d/2-s+1}^{(d-1)}(1)$.
For $s=1$ this  gives
\begin{equation}
  \gamma_1^{(d)}(\ell) \simeq \alpha_{-1/2}^{(d-1)}(1)+2-\frac{2}{\ell} 
  =\alpha_{d/2}^{(d-1)}(1)+2-\frac{2}{\ell} \,, 
  \label{gamma1_rel}
\end{equation}
and for $s>3/2$ 
\begin{equation}
  \gamma_s^{(d)}(\ell) \simeq \alpha_{d/2-s+1}^{(d-1)}(1)-\frac{2}{2s-3}
  + 4 \pi^{1-s} \Gamma(s) \zeta(2s-2) \ell^{2s-3} \,. 
  \label{gammas_rel}
\end{equation}
Note since $\zeta(0)=-1/2$ this also reproduces \eqref{gamma1_rel} for $s=1$.

For $d=4$ the above relations imply for $s=1$:
\begin{equation}
  \frac12 \left( \gamma_1^{(4)}(\ell)-\frac12\right) + \frac{1}{\ell}
  \simeq \rho \equiv 8\pi^2 \beta_2^{(3)}(1) 
  = \frac12 \alpha_2^{(3)}(1)+\frac34 \,,
  \label{rho_rel}
\end{equation}
for $s=2$:
\begin{equation}
 -\frac{1}{4\pi}\left( \gamma_2^{(4)}(\ell)-1\right) + \frac{\ell}{6}
 \simeq \beta_1^{(3)}(1) \,,
 \label{gamma2_rel}
\end{equation}
and for $s=3$:
\begin{equation}
  \gamma_3^{(4)}(\ell) - \frac{4\pi^2}{45} \ell^3
  \simeq \alpha_0^{(3)}(1) -\frac23\,. 
 \label{gamma3_rel}
\end{equation}

\subsection{The sunset diagram for large $\ell$ }
\label{sunset_ell}

The sunset diagram for the susceptibility is 
(cf. \cite{Niedermayer:2010mx} (4.1) and (4.36))
\begin{equation}
  \begin{split}
    \Psi(\ell,\ellhat) & = \Ls^{2D-4} \int_{V_D} \mrd x\, \ddot{G}(x) G^2(x)\\
    &= - \frac{1}{48\pi^2(D-4)}\left( 10\, \ddot{g}(0;\ell,\ellhat)
      - \frac{1}{\mcVD} \right) -\frac{1}{16\pi^2}\mcWov(\ell) 
    + \mrO(D-4)\,,
  \end{split}
  \label{Psi_d4}
\end{equation}
here we have reinstated $\ellhat$, but since we are working with 
$\ellhat=1$ we have $\mcVD=\ell\ellhat^{D-4}=\ell$.

For the analogous diagram in the infinite strip one had 
(cf. \cite{Niedermayer:2010mx} (5.11) and (5.61))
\begin{equation}
    \overline{\Psi}(\ellhat) = \Ls^{2D-4} \int_{V_\infty} \mrd x 
    \,\ddot{R}(x) R^2(x)
    = - \frac{1}{48\pi^2(D-4)} 10\, \ddot{R}(0;\ellhat) - c_w  + \mrO(D-4)\,.
  \label{ovPsi_d4}
\end{equation}
Further, for $\ell\gg 1$ one has up to exponentially small corrections
(cf. \cite{Niedermayer:2010mx}(5.20))
\begin{equation}
  \ddot{g}(0;\ell,\ellhat)- \frac{1}{\mcVD} \simeq \ddot{R}(0;\ellhat)\,.
\end{equation}
Since $\mcR(x) \simeq R(x)$ for $\ell\gg 1$ we have 
\begin{equation}
  \begin{split}
    \Delta\Psi &\equiv -\Ls^{2D-4}\int_{V_D} \mrd x
    \left[ \ddot{G}(x) G^2(x) - \ddot{\mcR}(x) \mcR^2(x) \right]
    \simeq -\Psi(\ell,\ellhat) + \overline{\Psi}(\ellhat) \\
    & = \frac{3}{16\pi^2 (D-4) \mcVD} 
    +\frac{1}{16\pi^2} \mcWov(\ell) - c_w\,.
  \end{split}
  \label{DPsi}
\end{equation}
Writing $G(x) = G_1(x) + \mcR(x)$ we have (noting $\int\mrd\bfx\,R(x)=0$)
\begin{equation}
   \Delta\Psi = 
   -\Ls^{2D-4} \int_{V_D} \mrd x \, \left\{ \ddot{G}_1(x) G^2(x) 
      + 2 G_1(x) \ddot{\mcR}(x) \mcR(x)\right \}
    = \Delta\Psi_1 + \Delta\Psi_2\,.
\end{equation}
The first term here is
\begin{equation}
  \begin{split}
    \Delta\Psi_1 & = 
    -\Ls^{2D-4} \int_{V_D} \mrd x \, \ddot{G}_1(x) G^2(x) = 
    \Delta\Psi_{1a} + \Delta\Psi_{1b}
    \\
    & = \Ls^{D-3}  \int_{\VdD}\mrd \bfx\, G^2(0,\bfx) 
    - \frac{1}{\ell}\Ls^{D-4}\int_{V_D} \mrd x \, G^2(x)\,.
  \end{split}
\end{equation}

\subsubsection{Calculating $\Delta\Psi_{1}$}

Since $\Delta\Psi_{1a}$ is regular at $D=4$ we can calculate it 
in 4-dimensions. 
Using \eqref{intddSS} one has
\begin{equation}
  \begin{split}
    & \Delta\Psi_{1a} = \Ls \int_{V_s} \mrd \bfx\, G^2(0,\bfx) 
    \\
    & = \frac{1}{16\pi^2 \Ls^3} \int_{V_s} \mrd \bfx
    \int_0^\infty \mrd u \mrd v\, \left\{
      (uv)^{-2}  S\left(\frac{\ell^2}{u}\right)
      S\left(\frac{\ell^2}{v}\right)
      \prod_{\mu=1}^{3} S\left(\frac{1}{u},\frac{x_\mu}{\Ls}\right)
      S\left(\frac{1}{v},\frac{x_\mu}{\Ls}\right)
    \right.
    \\
    & \qquad  -\frac{1}{\ell} u^{-2}S\left(\frac{\ell^2}{u}\right)
    \prod_{\mu=1}^{3} S\left(\frac{1}{u},\frac{x_\mu}{\Ls}\right)
    \left. 
      -\frac{1}{\ell} v^{-2}S\left(\frac{\ell^2}{v}\right)
      \prod_{\mu=1}^{3} S\left(\frac{1}{v},\frac{x_\mu}{\Ls}\right) 
      + \frac{1}{\ell^2}
    \right\}
    \\
    & = \frac{1}{16\pi^2} \int_0^\infty \mrd u \mrd v\, 
    (uv)^{-1/2}S\left(\frac{\ell^2}{u}\right)S\left(\frac{\ell^2}{v}\right)
    \big[ S^{3}(u+v)-1\big]
    \\
    & \qquad
    +\frac{\ell^2}{16\pi^2} 
    \left\{\int_0^\infty \mrd u\left(S(u)-1\right) \right\}^2
    \\
    & = \frac{1}{16\pi^2} \int_0^\infty \mrd u \mrd v\, 
    (uv)^{-1/2}\big[ S^{3}(u+v)-1\big] 
    + \frac{\ell^2}{16\pi^2}\left( \frac{2}{\pi}\zeta(2)\right)^2
    \\
    & = \frac{1}{16\pi}\left(
      \alpha_1^{(3)}(1) - 3\right) + \frac{\ell^2}{144} \,.
  \end{split}
\end{equation}

From \cite{Niedermayer:2010mx} (3.74),(3.30) one has
\begin{equation}
  \begin{split}
   \Delta\Psi_{1b} &= -\frac{1}{\mcVD} \Ls^{D-4}  \int_{V_D} G^2(x) 
   = \frac{1}{8\pi^2(D-4)\mcVD}
    - \frac{1}{16\pi^2\ell}\left(\alpha_2^{(4)}(\ell)-\frac{1}{2\ell}\right)
    \\
    & \simeq \frac{1}{8\pi^2(D-4)\mcVD} - \frac{1}{720}\ell^2
    - \frac{1}{16\pi^2\ell}\left(\alpha_0^{(3)}(1)-\frac23\right)\,.
  \end{split}
\end{equation}

Combining the results we get
\begin{equation}
  \Delta\Psi_1 \simeq \frac{1}{8\pi^2(D-4)\mcVD} 
  -\frac{1}{16\pi^2\ell}\left(\alpha_0^{(3)}(1)-\frac23\right)
  +\frac{1}{16\pi}\left(\alpha_1^{(3)}(1)-3\right) + \frac{\ell^2}{180}\,.
\end{equation}

\subsubsection{Calculating $\Delta\Psi_2$}

\begin{equation}
  \begin{split}
    \Delta\Psi_{2} &= 
    -2 \Ls^{2D-4} \int_{V_D} \mrd x \, G_1(x) \ddot{\mcR}(x) \mcR(x)
    = \Delta\Psi_{2a} + \Delta\Psi_{2b}
    \\
    & = -2 \Ls^{2D-4} \int_{V_D} \mrd x \, G_1(x) \ddot{G}(x) G(x)
    + 2 \Ls^{2D-4} \int_{V_D} \mrd x \, \ddot{G}_1(x) G_1^2(x)\,.
  \end{split}
  \label{DeltaPsi2}
\end{equation}

We have
\begin{equation}
  \begin{split}
   &\Delta\Psi_{2a} = -2 \Ls^{2D-4} \int_{V_D} \mrd x \, G_1(x) \ddot{G}(x) G(x)
   = -2 \frac{1}{16\pi^2} \int_{-1/2}^{1/2} \mrd z_0\,\Delta_1(z_0) 
   \\
   & \qquad
   \times \int_0^\infty \mrd u \mrd v\, 
   u^{-1/2} \ddot{S}\left(\frac{\ell^2}{u},z_0\right)
   \left\{ v^{-1/2} S\left(\frac{\ell^2}{v},z_0\right)S^{D-1}(u+v)
   -\frac{1}{\ell} \right\}
   \\
   & = -2 \frac{1}{16\pi^2} \int_{-1/2}^{1/2} \mrd z_0\,\Delta_1(z_0)
   \\
   & \qquad
   \times \int_0^\infty \mrd u  \mrd v\, \Bigg\{
   (uv)^{-1/2} \ddot{S}\left(\frac{\ell^2}{u},z_0\right)
   S\left(\frac{\ell^2}{v},z_0\right)\left[S^{D-1}(u+v)-1\right]
   \\
   & \qquad\qquad
   + u^{-1/2} \ddot{S}\left(\frac{\ell^2}{u},z_0\right) 
   \left[u^{-1/2}S\left(\frac{\ell^2}{v},z_0\right)-\frac{1}{\ell}\right]
   \Bigg\}
   \\
   & \simeq -2 \frac{1}{16\pi^2} \int_{-1/2}^{1/2} \mrd z_0 
   \Delta_1(z_0)
   \Bigg\{ \int_0^\infty \mrd u  \mrd v\, 
   (uv)^{-1/2} \left(-2\pi\ell^2 u^{-1} + 4\pi^2\ell^4 u^{-2} z_0^2 \right)
   \\
   & \qquad
   \times 
   \mre^{-\pi z_0^2 \ell^2 (1/u+1/v)}\left[S^{D-1}(u+v)-1\right]
   \Bigg\} 
   - 2\ell^2 \int_{-1/2}^{1/2}\mrd z_0\,\ddot{\Delta}_1(z_0)\Delta_1^2(z_0)\,. 
  \end{split}
\end{equation}
The second integral cancels with $\Delta\Psi_{2b}$ in \eqref{DeltaPsi2}.

For $\ell^2(1/u+1/v) \gg 1$ one has
\begin{equation}
  \int_{-1/2}^{1/2} |z_0|^r \exp\left(-\pi z_0^2 \ell^2(1/u+1/v)\right)
  \simeq \frac{\Gamma\left(\frac{r+1}{2}\right)}{\pi^{(r+1)/2} \ell^{r+1}}
  \left(\frac{uv}{u+v} \right)^{(r+1)/2}\,.
\end{equation}

Integrating over $z_0$ one obtains 
\footnote{
Note $\int_0^\infty \mrd u  \mrd v\, f(u+v) g(u,v) =
  \frac12 \int_0^\infty \mrd t\, f(t) \int_{-t}^t \mrd \eta\,
  g((t+\eta)/2,(t-\eta)/2) \,.$
The integral over $\eta$ in our case gives a power of $t$.}
\begin{equation}
  \begin{split}
   \Delta\Psi_2 & \simeq -2\frac{1}{16\pi^2}
   \int_0^\infty \mrd u  \mrd v\, \left[S^{D-1}(u+v)-1\right]
   \\
   & \qquad  
   \times \left[
     -\frac{\pi}{6 (u+v)^{3/2}} \ell
     + \frac{\sqrt{v}(u-v)}{\sqrt{u}(u+v)^2}
     -\frac{v(u-2v)}{2(u+v)^{5/2}} \frac{1}{\ell}\right]
   \\
   & = \frac{1}{16\pi^2}\left[
   \frac{\pi}{3}\left(\alpha_1^{(3)}(1)-3\right)\ell
   +\frac{\pi}{2}\left(\alpha_1^{(3)}(1)-3\right) \right.
   \\
   & \qquad \left.
     +\frac{1}{(D-4)\mcVD}
   -\frac12 \left( \alpha_{0}^{(3)}(1)-\frac23\right)\frac{1}{\ell}
   \right]\,.
 \end{split}
\end{equation}

Finally using \eqref{beta_k} one gets
\begin{equation}
  \begin{split}
    \Delta\Psi & \simeq
    \frac{3}{16\pi^2(D-4)\mcVD} 
    -\frac{1}{12} \beta_1^{(3)}(1) \left(\ell+\frac92\right) 
    \\
    & \quad
    -\frac{1}{16\pi^2} 
    \left( \frac32 \alpha_0^{(3)}(1)-1\right)\frac{1}{\ell}
    +\frac{\ell^2}{180}\,.
  \end{split}
  \label{mPsi12}
\end{equation}
Comparing this with \eqref{DPsi} one obtains the desired relation 
\begin{equation}
  \frac{1}{16\pi^2} \mcWov(\ell) - \frac{1}{180} \ell^2 
  +\frac{1}{12} \beta_1^{(3)}(1) \left(\ell+\frac92\right)
  +\frac{1}{16\pi^2} 
    \left( \frac32 \alpha_0^{(3)}(1)-1\right)\frac{1}{\ell}
    \simeq c_w\,.
  \label{R2T2rel}
\end{equation}
The lhs converges exponentially; e.g. for   
$\ell=4$ it is evaluated as $0.0986829793$ which agrees
to 8 significant figures with \eqref{cw}.
In fact already at $\ell=2$ the lhs is very close to $c_w$ 
as one can check using the result eq.(4.45) in \cite{Niedermayer:2010mx}.

\section{Details of the perturbative perturbative
computation in subsect.~3.1}
\label{Apppertdetails}

\subsection{Contributions to the action}
\label{action_contribs}

The four derivative terms in \eqref{A4total_cont}
are given by ($\int_x\dots=\int_{V_D}\mrd x\dots$):
\begin{align}
  &A_4^{(2)}=\int_x\sum_{\mu\nu} (\pmu\bS(x)\pmu\bS(x))
  (\pnu\bS(x)\pnu\bS(x))\,,
  \label{A42cont}\\
  &A_4^{(3)}=\int_x\sum_{\mu\nu} (\pmu\bS(x)\pnu\bS(x))
  (\pmu\bS(x)\pnu\bS(x))\,.
  \label{A43cont}
\end{align}

Terms appearing in \eqref{A2hA4h} are given by the following:
\begin{align}  \label{jmu}
  B_2 & = -\frac{1}{g_0^2} \int_x\,j_0(x)\,,\,\,\,\,\,\,
  j_\mu(x)=S_2(x)\partial_\mu S_1(x)-S_1(x)\partial_\mu S_2(x)\,,\\
  C_2 & = \frac{1}{2g_0^2}\int_x [Q\bS(x)]^2\,.
\end{align}
For the operator 2:
\begin{align}
  B_4^{(2)}&=-4\int_x (\pmu\bS(x)\pmu\bS(x)\,j_0(x)\,,
  \\
  C_4^{(2)}&=-2\int_x\left\{(\pmu\bS(x)\pmu\bS(x))
    \left[S_1(x)^2+S_2(x)^2\right]+2\left[j_0(x)\right]^2\right\}\,,
\end{align}
and for the operator 3:
\begin{align}
  B_4^{(3)}&=-4\int_x(\partial_0\bS(x)\pmu\bS(x))\,j_\mu(x)\,,
  \\
  C_4^{(3)}&=-2\int_x\left\{(\partial_0\bS(x)\partial_0\bS(x))
    \left[S_1(x)^2+S_2(x)^2\right]
    +2\left[j_0(x)\right]^2+\left[j_k(x)\right]^2\right\}\,.
\end{align}

\subsection{Averages over rotations}
\label{rot_averages}

In this section we set $\bS(x) = \Omega \bR(x)$ in the
correlation functions in \eqref{fhx}
and average over the rotations.

For the mass independent terms we have 
\begin{equation} \label{Ah1}
  \langle A_{2h} \rangle =  -\frac{h^2 V_D}{n g_0^2}\,,
\end{equation}
and (using \eqref{avabcd2a}):
\begin{equation} \label{Ah2}
  -\frac12  \langle A_{2h}^2 \rangle  
=\frac{2 h^2V_D}{n(n-1) g_0^4}\langle W\rangle\,,
\end{equation}
where 
\begin{equation}\label{W}
  W = \frac{1}{V_D} \int_{x y}
    \left( \partial_0 \bR(x) \partial_0 \bR(y) \right)
    \Bigl[\left( \bR(x) \bR(y) \right)-1\Bigr]\,. 
\end{equation}
For the 4-derivative terms:
\begin{align} \label{C42av}
  \langle C_4^{(2)}\rangle &=-\frac{4}{n(n-1)}\int_x
  \langle (n-1)(\partial_\mu\bR(x)\partial_\mu\bR(x))
    +2(\partial_0\bR(x)\partial_0\bR(x))\rangle\,,
  \\ \label{C43av}
  \langle C_4^{(3)}\rangle &=-\frac{4}{n(n-1)}\int_x
  \langle      (\partial_\mu\bR(x)\partial_\mu\bR(x))
    +n(\partial_0\bR(x)\partial_0\bR(x))\rangle\,.
\end{align}

Turning now to the mass terms we get
\begin{equation} \label{Am41}
  \langle A_{M;4}^{(1)} \rangle
  = \frac{1}{n} M^4 V_D \,,
\end{equation}
and
\begin{align} \label{Am2}
  \langle A_{M;2}^2 \rangle
&=\frac{M^4}{ng_0^4}\int_{yz}
\langle(\bR(y)\bR(z))\rangle
\\
&=\frac{M^4V_D^2}{ng_0^4}\left[1+P_1\right]\,,
\end{align}
where
\begin{equation} \label{P1}
P_1=\frac{1}{V_D^2}\int_{yz}\langle(\bR(y)\bR(z))-1\rangle\,.
\end{equation}

Using \eqref{avabcd2}:
\begin{align} 
\langle C_2 A_{M;2}^2 \rangle&=-\frac{M^4}{g_0^6}
\int_{xyz}\langle S_1(x)^2S_n(y)S_n(z)\rangle
\\ 
&=-\frac{M^4}{g_0^6}\frac{1}{n(n-1)(n+2)}\times
\nonumber\\ \label{AhAm2} 
&\times\int_{yz}
\langle (n+1)V_D(\bR(y)\bR(z))
-2\int_x(\bR(x)\bR(y))(\bR(x)\bR(z))\rangle
\\
&=-\frac{M^4V_D^3}{g_0^6}\frac{1}{n(n-1)(n+2)}
\left[n-1+(n-3)P_1-2P_2\right]\,,
\end{align}
where
\begin{equation}\label{P2}
P_2=\frac{1}{V_D^3}\int_{xyz}
\langle\left[(\bR(x)\bR(y))-1\right]\left[(\bR(x)\bR(z))-1\right]\rangle\,.
\end{equation}

Next, using \eqref{avabcdef2}:
\begin{align}
\langle B_2^2 A_{M;2}^2 \rangle&=\frac{M^4}{g_0^8}
\int_{wxyz}\langle j_0(w)j_0(x)S_n(y)S_n(z)\rangle
\\
&=\frac{2M^4V_D^3}{n(n-1)(n-2)(n+2)g_0^6}P_3\,,
\end{align}
where 
\begin{align}
P_3&=\frac{1}{V_D^3g_0^2}\int_{wxyz}
\nonumber\\
& \langle n(\bR(y)\bR(z))\left[(\bR(w)\bR(x))(\partial_0\bR(w)\partial_0\bR(x))
-(\bR(w)\partial_0\bR(x))(\bR(x)\partial_0\bR(w))\right]    
\nonumber\\
& + 2(\bR(x)\bR(z))\left[
(\bR(w)\partial_0\bR(x))(\bR(y)\partial_0\bR(w))
    -(\partial_0\bR(w)\partial_0\bR(x))(\bR(w)\bR(y))\right]
\nonumber\\ 
& +2(\bR(z)\partial_0\bR(x))\left[
(\bR(x)\partial_0\bR(w))(\bR(w)\bR(y))
-(\bR(w)\bR(x))(\bR(y)\partial_0\bR(w))\right]\rangle\,.
\end{align}
By partial integration this simplifies to
\begin{align}\label{P3}
P_3&=\frac{2}{V_D^3g_0^2}\int_{wxyz}
\nonumber\\
&\langle \left[n(\bR(y)\bR(z))(\bR(w)\bR(x))
-4(\bR(x)\bR(z))(\bR(w)\bR(y))\right](\partial_0\bR(w)\partial_0\bR(x))
\rangle
\\
&= \frac{2n}{g_0^2}\langle W\rangle+\overline{P}_3\,,
\end{align}
where $W$ is given in \eqref{W} and
\begin{align}
\overline{P}_3&=\frac{2}{V_D^3g_0^2}\int_{wxyz}
\nonumber\\
&\langle \left[n\{(\bR(y)\bR(z))-1\}\{(\bR(w)\bR(x))-1\}
-4\{(\bR(x)\bR(z))-1\}\{(\bR(w)\bR(y))-1\}
\right]
\nonumber\\
&\times(\partial_0\bR(w)\partial_0\bR(x))\rangle\,.
\end{align}

Next
\begin{align} \label{Am2Am42}
\langle A_{M;2}A_{M;4}^{(2)}\rangle
&=-\frac{M^4}{2g_0^2}\int_{xy} S_n(x)S_n(y)(\pmu \bS(y) \pmu \bS(y))
\\
&=-\frac{M^4V_D^2}{2ng_0^2} P_4\,,
\end{align}
where
\begin{equation} \label{P4}
P_4=\frac{1}{V_D^2}\int_{xy} (\bR(x)\bR(y))(\pmu\bR(y)\pmu\bR(y))\,,
\end{equation}
and
\begin{align}
\langle C_2 A_{M;4}^{(1)}\rangle
&=-\frac{M^4}{g_0^2}\int_{xy}\langle S_1(x)^2S_n(y)^2\rangle
\\
&=-\frac{M^4}{g_0^2}\frac{1}{n(n-1)(n+2)}
\int_y\langle (n+1)V_D-2\int_x(\bR(x)\bR(y))^2\rangle
\\
&=-\frac{M^4V_D^2}{g_0^2}\frac{1}{n(n-1)(n+2)}
\left[n-1+(n-3)P_1-2P_5\right]\,,
\end{align}
where 
\begin{equation}
P_5=\frac{1}{V_D^2}\int_{xy}\langle\left[(\bR(x)\bR(y))-1\right]^2\rangle
=\mrO(g_0^4)\,.
\end{equation}

Proceeding to the terms $\langle C_4^{(i)}A_{M;2}^2\rangle\,,$ first
\begin{equation}
\langle C_4^{(2)}A_{M;2}^2\rangle=
-\frac{4M^4}{g_0^4}\left[Q^{(2A)}+Q^{(2B)}\right]\,,
\end{equation}
with
\begin{align}
Q^{(2A)}&=\int_{xyz}\langle 
(\pmu\bS(x)\pmu\bS(x))S_1(x)^2S_n(y)S_n(z)\rangle
\\
&=\frac{1}{n(n-1)(n+2)}\times
\nonumber\\ 
&\times\int_{xyz}
\langle (\pmu\bR(x)\pmu\bR(x))\left[(n+1)(\bR(y)\bR(z))
-2(\bR(x)\bR(y))(\bR(x)\bR(z))\right]\rangle
\\
&=\frac{V_D^2}{n(n+2)}\int_x
\langle (\pmu\bR(x)\pmu\bR(x))\rangle+\mrO(g_0^4)\,,
\end{align}
and
\begin{align}
Q^{(2B)}&=\int_{xyz}\langle j_0(x)^2 S_n(y)S_n(z)\rangle\,,
\\
&=\frac{2V_D^2}{n(n-1)(n-2)(n+2)}P_6\,,
\end{align}
where 
\begin{align}
P_6&=\frac{1}{V_D^2}\int_{xyz}
\langle n(\bR(y)\bR(z))(\partial_0\bR(x)\partial_0\bR(x))
\nonumber\\
& - 2(\bR(x)\bR(z))(\bR(x)\bR(y))(\partial_0\bR(x)\partial_0\bR(x))
-2(\bR(z)\partial_0\bR(x))(\bR(y)\partial_0\bR(x))\rangle
\\
&=(n-2)\int_x \langle (\partial_0\bR(x)\partial_0\bR(x))\rangle
+\mrO(g_0^4)\,. 
\end{align}
So
\begin{equation}
Q^{(2A)}+Q^{(2B)}=
-\frac{V_D^2}{4(n+2)}\langle C_4^{(2)}\rangle+\mrO(g_0^4)\,.
\end{equation}

Similarly
\begin{equation}
\langle C_4^{(3)}A_{M;2}^2\rangle=
-\frac{4M^4}{g_0^4}\left[Q^{(3A)}+Q^{(2B)}+Q^{(3C)}\right]\,,
\end{equation}
with
\begin{align}
Q^{(3A)}&=\int_{xyz}\langle 
(\partial_0\bS(x)\partial_0\bS(x))S_1(x)^2S_n(y)S_n(z)\rangle
\\
&=\frac{V_D^2}{n(n+2)}\int_x
\langle (\partial_0\bR(x)\partial_0\bR(x))\rangle+\mrO(g_0^4)\,,
\\
Q^{(3C)}&=\frac12\int_{xyz}\langle j_k(x)^2 S_n(y)S_n(z)\rangle
\\
&=\frac{V_D^2}{n(n-1)(n+2)}
\int_x \langle (\partial_k\bR(x)\partial_k\bR(x))\rangle+\mrO(g_0^4)\,. 
\end{align}
So
\begin{equation}
Q^{(3A)}+Q^{(3B)}+Q^{(3C)}=
-\frac{V_D^2}{4(n+2)}\langle C_4^{(3)}\rangle+\mrO(g_0^4)\,.
\end{equation}
Together we have simply
\begin{equation}
\langle C_4 A_{M;2}^2\rangle=
\frac{M^4V_D^2}{(n+2)g_0^4}\langle C_4\rangle+\mrO(M^4g_0^0)\,.
\end{equation}

The next two averages in \eqref{fhx} to be taken into account are
\begin{align}
\langle C_2 A_{M;2}A_{M;4}^{(2)}\rangle
&=\frac{M^4}{2g_0^4}\int_{xyz}\langle  S_1(x)^2 S_n(y)S_n(z)
(\pmu\bS(z)\pmu\bS(z))\rangle
\\
&=\frac{M^4}{2g_0^4}\frac{1}{n(n-1)(n+2)}\times
\nonumber\\
&\times\int_{xyz}\langle
(\pmu\bR(z)\pmu \bR(z))\left[(n+1)(\bR(y)\bR(z))
-2(\bR(x)\bR(y))(\bR(x)\bR(z))\right]\rangle
\\
&=\frac{M^4V_D^2}{2g_0^4}\frac{1}{n(n+2)}\int_z
\langle (\pmu\bR(z)\pmu\bR(z))\rangle+\mrO(M^4g_0^0)\,,
\end{align}
and
\begin{align}
\langle A_{M;2}C_{M;4} \rangle
&=\frac{M^4}{g_0^2}L_2\int_{xy}\langle S_n(y)S_n(x)S_1(x)^2 \rangle
\\
&=\frac{M^4}{n(n+2)g_0^2}L_2\int_{xy}\langle (\bR(x)\bR(y))\rangle
\\
&=\frac{M^4V_D^2}{n(n+2)g_0^2}L_2+\mrO(g_0^0)\,.
\end{align}

With similar considerations we can show that the last
4 terms in \eqref{fhx} can be neglected to the order of interest:
\begin{equation}
\langle B_2^2 A_{M;4}^{(1)}\rangle\,,
\langle B_2^2 A_{M;2}A_{M;4}^{(2)}\rangle\,,
\langle B_2B_4^{(2)} A_{M;2}^2\rangle\,,
\langle B_2B_4^{(3)} A_{M;2}^2\rangle
=\mrO(M^4g_0^0)\,.
\end{equation}

\subsection{Perturbative computations}
\label{pert_calc}

After separating the zero mode and changing to $\vpi$ variables
according to $\bR = (g_0 \vpi, \sqrt{1-g_0^2 \vpi^2})$ the
effective action is given by
\begin{equation}
  A_\mathrm{eff}[\vec{\pi}]=A[\vec{\pi}]+A_{\mathrm{zero}}[\vec{\pi}]\,,
\end{equation}
where with DR we have dropped the measure term, and
\begin{equation}
  A_{\mathrm{zero}}[\vpi]=
  -(n-1)\ln \left( 
    \frac{1}{V_D}\int_x \left( 1-g_0^2\vpi(x)^2\right)^\frac12 \right).
\end{equation}
The effective action has a perturbative expansion
\begin{equation} \label{Aeff}
  A_\mathrm{eff} = 
  A_0 + g_0^2 A_1 + g_0^4 A_2 + \ldots
\end{equation}
where
\begin{equation}
  \label{A0}
  A_0 = \frac12 \int_x\partial_\mu\vpi(x)\partial_\mu\vpi(x)\,,
\end{equation}
and
\begin{equation}\label{A1}
A_1=A_1^{(a)}+A_1^{(b)}\,,
\end{equation}
with
\begin{align}
  A_1^{(a)} & = \frac18\int_x\partial_\mu[\vpi(x)^2]\partial_\mu[\vpi(x)^2]\,,
\\
  A_1^{(b)} & = \frac12\frac{(n-1)}{V_D}\int_x\vpi(x)^2\,.
\end{align}

For the details of the perturbative computation of $\chi_0$
we refer the reader to \cite{Niedermayer:2016yll}.

$P_j$ have perturbative expansions (to the order we need)
\begin{equation} 
  P_j = P_j^{(1)}g_0^2+P_j^{(2)}g_0^4+\ldots\,,j=1,2,3\,.
\end{equation}
The expansion of $\chi_1$ is then
\begin{align}  
\chi_1&=-\frac{1}{2n}\left[1+P_1\right]\chi_0
+\frac{1}{n(n-1)(n+2)g_0^2}
\left[n-1+(n-3)P_1-2P_2-\frac{1}{(n-2)}P_3\right]
\nonumber\\
&-\frac{1}{(n+2)V_D}\langle C_4\rangle
+\frac{2g_0^2}{n^2(n+2)V_D}\left[2L_1+L_2\right]+\mrO(g_0^4)
\\
&=-\frac{2}{n^2(n+2)g_0^2}\left[1+\chi_1^{(1)}g_0^2
+\chi_1^{(2)}g_0^4+\dots\right]\,,
\end{align}
with    
\begin{align}
\chi_1^{(1)}&=\frac{1}{(n-1)}\left[(2n-1)P_1^{(1)}
+nP_2^{(1)}+\frac{n}{2(n-2)}\overline{P}_3^{(1)}
-\frac{2(n-1)}{(n-2)}R_1\right]\,,
\\
\chi_1^{(2)}&=\frac{1}{(n-1)}\left[(2n-1)P_1^{(2)}
+nP_2^{(2)}+\frac{n}{2(n-2)}\overline{P}_3^{(2)}
-\frac{2(n-1)}{(n-2)}R_2^{(a)}\right]
\nonumber\\
&-\frac{1}{2n}R_1P_1^{(1)}+R_2^{(b)}-\frac{1}{V_D}\left[2L_1+L_2\right]\,.
\end{align}
Here $R_1\,,R_2^{(a)},R_2^{(b)}$ appear in the 
perturbative expansions of $\langle W \rangle$ and $\langle C_4 \rangle$
\cite{Niedermayer:2016yll}: 
\begin{align}
\frac{1}{g_0^2}\langle W \rangle
&=-\frac12 (n-1)\left[g_0^2 R_1 + g_0^4 R_2^{(a)}+\dots\right]\,,
\\
\langle C_4 \rangle&=-\frac{V_D}{n}g_0^2 R_2^{(b)} + \ldots\,.
\end{align}

In the following we will need the expansions
\begin{align} \label{RR}
  \left( \bR(x) \bR(y) \right) &= 1  
  - \frac12 g_0^2 [\vpi(x)-\vpi(y)]^2 
  -\frac18 g_0^4 [\vpi(x)^2-\vpi(y)^2]^2 
  + \ldots\,,
\\ \label{dRdR}
  \left( \partial_0\bR(x) \partial_0\bR(y) \right)&= 
  g_0^2 \left( \partial_0\vpi(x)\partial_0\vpi(y) \right)  
  +\frac14 g_0^4 \partial_0[\vpi(x)^2]\partial_0[\vpi(y)^2]+\dots\,.
\end{align}

\subsubsection{NL order}

From \eqref{RR} we have
\begin{align} 
P_1^{(1)}&=-\frac{1}{V_D^2}\int_{yz}\langle \vpi(y)^2\rangle_0
\\
&=-(n-1)\IDR_{10}\,.
\end{align}
Here $\IDR_{10}$ is a particular case of the dimensionally regularized sums
\begin{equation}
  \IDR_{nm}\equiv\frac{1}{V_D}\psump\frac{p_0^{2m}}{(p^2)^n}\,,
\end{equation}
which are discussed in \cite{Niedermayer:2010mx}.

There are no contributions from $P_2,\overline{P}_3$ to the NL order:
\begin{equation}
P_2^{(1)}=0=\overline{P}_3^{(1)}\,.
\end{equation}
So
\begin{equation}
\chi_1^{(1)}=-(2n-1)\IDR_{10}+4\IDR_{21}\,.
\end{equation}
Recalling for $D=4$:
\begin{align}
\IDR_{10}&=-\frac{1}{\Ls^2}\beta_1\,,
\\
\IDR_{21}&=\frac{1}{8\pi \Ls^2}\left(\gamma_2-1\right)\,,
\end{align}
we obtain \eqref{chi11_pert}.

\subsubsection{NNL order}

First we have
\begin{equation}
P_1^{(2)}=P_1^{(2a)}+P_1^{(2b)}+P_1^{(2c)}\,,
\end{equation}
with
\begin{align}
P_1^{(2a)}&=\frac{1}{V_D}\int_y\langle\vpi(y)^2A_1^{(a)}\rangle_0^c
\\
&=\frac{(n-1)}{V_D}\int_{xy}\left\{\partial_\mu^z\partial_\mu^x
\left[G(y-z)G(y-x)G(x-z)\right]\right\}_{z=x}
\\
&=(n-1)\left[\IDR_{10}^2-\frac{1}{V_D}\IDR_{20}\right]\,,
\\
P_1^{(2b)}&=\frac{1}{V_D}\int_y\langle\vpi(y)^2A_1^{(b)}\rangle_0^c
\\
&=(n-1)^2\frac{1}{V_D}\IDR_{20}\,,
\\
P_1^{(2c)}&=-\frac{1}{8V_D^2}\int_{yz}\langle[\vpi(y)^2- \vpi(z)^2]^2\rangle_0
\\
&=-\frac12 (n-1)\left[\IDR_{10}^2-\frac{1}{V_D}\IDR_{20}\right]\,.
\end{align}
So
\begin{equation}
P_1^{(2)}=\frac12(n-1)\left[\IDR_{10}^2+(2n-3)\frac{1}{V_D}\IDR_{20}\right]\,.
\end{equation}

Next
\begin{align}
P_2^{(2)}&=\frac{1}{4V_D^3}\int_{xyz}
\langle[\vpi(x)-\vpi(y)]^2[\vpi(x)-\vpi(z)]^2\rangle_0
\\
&=\frac12(n-1)\left[(2n-1)\IDR_{10}^2+\frac{3}{V_D}\IDR_{20}\right]\,.
\end{align}
 
Finally
\begin{align}
\overline{P}_3^{(2)}&=-\frac{2}{V_D^2}\int_{wxy}
\langle \left[n\vpi(y)^2(\vpi(w)\vpi(x))+\vpi(w)^2\vpi(x)^2\right]
(\partial_0\vpi(w)\partial_0\vpi(x))\rangle_0
\\
&=-2n(n-1)(n-2)\left[(n-1)\IDR_{10}\IDR_{21}+\frac{4}{V_D}\IDR_{31}\right]
-4(n-1)(n-2)\Wov\,,
\end{align}
where we have used
\begin{equation}
\int_x G(x)\partial_0 G(x)\partial_0 G(x)=\frac12\Wov\,. 
\end{equation}

So we obtain
\begin{align}
\chi_1^{(2)}&=R_2+\frac12(2n-1)(n+1)\IDR_{10}^2
+ 4n\IDR_{21}^2-4(2n-1)\IDR_{10}\IDR_{21}
\nonumber\\
&+\frac{1}{V_D}\left[-8n(n-1)\IDR_{31}
+\frac12(4n^2-5n+3)\IDR_{20}-2L_1-L_2\right]\,.
\end{align}                                         
Note the result depends only on the combination $2L_1+L_2$
as anticipated from our discussion in subsect.~\ref{PTmass}.

Recalling for $D\sim 4$ 
\begin{align}
  \IDR_{20}&=\frac{1}{8\pi^2}\left[\ln \Ls-\frac{1}{D-4}
  +\frac12\alpha_2-\frac{1}{4\ell}\right]+\mrO(D-4)\,,
  \\
  \IDR_{31}&=\frac{1}{32\pi^2}\left[\ln \Ls-\frac{1}{D-4}
  +\frac12\gamma_3\right]+\mrO(D-4)\,,
\end{align}
we see that to cancel the pole terms the combination 
$L_1+\frac12 L_2$ must be of the form given in \eqref{L1L2}.
Finally combining the results we obtain \eqref{chi12_pert}.

\section{Integrals over O($n$) matrices}
\label{AppHaar}

Here we consider integrals of the form
\begin{equation}
H^{(r)}_{i_1\dots i_{2r};j_1\dots j_{2r}}
\int\mrd\Omega\,\prod_{a=1}^{2r}\Omega_{i_aj_a}\,,
\end{equation} 
with $\Omega$ a real orthogonal $n\times n$ matrix:
\begin{equation}
\sum_{k=1}^n\Omega_{ik}\Omega_{jk}=\delta_{ij}
=\sum_{k=1}^n\Omega_{ki}\Omega_{kj}\,,
\end{equation} 
and $\mrd\Omega$ the O($n$) invariant Haar measure with normalization
\begin{equation}
\int\mrd\Omega = 1\,.
\end{equation}

For $r\ge1$ the integrals are evaluated as sums over $r$ terms: 
\begin{equation}\label{Hr}
H^{(r)}_{i_1\dots i_{2r};j_1\dots j_{2r}}
=\sum_{s=1}^r\,
\left\{\left[\prod_{b=1}^r\delta_{i_{2b-1}i_{2b}}\right]
T^{(r)(s)}_{j_1\dots j_{2r}}+(N_r-1)\,\,{\rm perms}\right\}\,,
\end{equation} 
where the  $T^{(r)(s)}$ are $n$-dependent coefficients times
sums of products of $\delta$-functions
involving only the $j_a$ which are invariant under 
interchanges $j_a\to j_{\sigma(a)}\,,\,\,a=1,..,2r$,
where $\sigma$ is any permutation such that
\begin{equation}
\prod_{b=1}^r\delta_{i_{2b-1}i_{2b}}
=\prod_{b=1}^r\delta_{i_{\sigma(2b-1)}i_{\sigma(2b)}}\,.
\end{equation} 
The sum over permutations in \eqref{Hr} gives the sum over the 
$N_r=(2r)!/(2^r r!)$ distinct pairings of the labels $i_1,\dots,i_{2r}$. 

In our computation we only need the integrals for $r=1,2,3$.
First for $r=1$:
\begin{equation}
H^{(1)}_{i_1i_2;j_1j_2}=\frac{1}{n}\delta_{i_1i_2}\delta_{j_1j_2}\,,
\end{equation}
For $r=2$ and $n>1$:
\begin{align}
T^{(2)(1)}_{j_1j_2j_3j_4}&=\frac{(n+1)}{n(n-1)(n+2)}
\delta_{j_1j_2}\delta_{j_3j_4}\,,
\\
T^{(2)(2)}_{j_1j_2j_3j_4}&=-\frac{1}{n(n-1)(n+2)}\left[
\delta_{j_1j_3}\delta_{j_2j_4}+\delta_{j_1j_4}\delta_{j_2j_3}\right]\,.
\end{align}
Finally for $r=3$ and $n>2$:
\begin{align}
T^{(3)(1)}_{j_1j_2j_3j_4j_5j_6}&=
\frac{(n^2+3n-2)}{n(n-1)(n-2)(n+2)(n+4)}
\delta_{j_1j_2}\delta_{j_3j_4}\delta_{j_5j_6}\,,
\\
T^{(3)(2)}_{j_1j_2j_3j_4j_5j_6}&=
-\frac{1}{n(n-1)(n-2)(n+4)}\left\{
\delta_{j_1j_2}\left[\delta_{j_3j_5}\delta_{j_4j_6}
                    +\delta_{j_3j_6}\delta_{j_4j_5}\right]\right.
\nonumber\\
&\left.+\delta_{j_3j_4}\left[\delta_{j_1j_5}\delta_{j_2j_6}
                            +\delta_{j_1j_6}\delta_{j_2j_5}\right]
       +\delta_{j_5j_6}\left[\delta_{j_1j_3}\delta_{j_2j_4}
                            +\delta_{j_1j_4}\delta_{j_2j_3}\right]\right\}\,,
\\
T^{(3)(3)}_{j_1j_2j_3j_4j_5j_6}&=
\frac{2}{n(n-1)(n-2)(n+2)(n+4)}\times
\nonumber\\
&\times\left\{
 \delta_{j_1j_3}\left[\delta_{j_2j_5}\delta_{j_4j_6}
                     +\delta_{j_2j_6}\delta_{j_4j_5}\right]
+\delta_{j_1j_4}\left[\delta_{j_2j_5}\delta_{j_3j_6}
                     +\delta_{j_2j_6}\delta_{j_3j_5}\right]\right.
\nonumber\\
&+\left.
 \delta_{j_1j_5}\left[\delta_{j_2j_3}\delta_{j_4j_6}
                     +\delta_{j_2j_4}\delta_{j_3j_6}\right]
+\delta_{j_1j_6}\left[\delta_{j_2j_3}\delta_{j_4j_5}
                     +\delta_{j_2j_4}\delta_{j_3j_5}\right]\right\}\,.
\end{align}

It follows for averages over O$(n)$ vectors $a,b,c,d,e,f$:
\begin{align}
&\langle a_1b_1\rangle = \frac{1}{n}\langle a\cdot b\rangle\,,
\\ \label{avabcd2}
&\langle a_1b_1c_2d_2\rangle = \frac{1}{n(n-1)(n+2)}
\langle (n+1)(a\cdot b)(c\cdot d)
-(a\cdot c)(b\cdot d)-(a\cdot d)(b\cdot c)\rangle\,,
\\ \label{avabcd2a}
&\langle(a_1b_2-a_2b_1)(c_1d_2-c_2d_1)\rangle =  
\frac{2}{n(n-1)}\langle(a\cdot c)(b\cdot d)-(a\cdot d)(b\cdot c)\rangle\,,
\\
&\langle a_1b_1c_2d_2e_3f_3\rangle 
=\frac{1}{n(n-1)(n-2)(n+2)(n+4)}\langle
(n^2+3n-2)(a\cdot b)(c\cdot d)(e\cdot f) 
\nonumber\\
&-(n+2)\left((a\cdot b)\left[(c\cdot e)(d\cdot f)
+(c\cdot f)(d\cdot e)\right]\right.
\nonumber\\
&\left.+(c\cdot d)\left[(a\cdot e)(b\cdot f)        
+(a\cdot f)(b\cdot e)\right]
+(e\cdot f)\left[(a\cdot c)(b\cdot d)        
+(a\cdot d)(b\cdot c)\right]\right)
\nonumber\\
&+2\left( 
 (a\cdot c)\left[(b\cdot e)(d\cdot f)
+(b\cdot f)(d\cdot e)\right]
+(a\cdot d)\left[(b\cdot e)(c\cdot f)
+(b\cdot f)(c\cdot e)\right]\right.
\nonumber\\
&+\left.
 (a\cdot e)\left[(b\cdot c)(d\cdot f)
+(b\cdot d)(c\cdot f)\right]
+(a\cdot f)\left[(b\cdot c)(d\cdot e)
+(b\cdot d)(c\cdot e)\right]\right)\rangle\,,
\\
&\langle \left(a_1b_2-a_2b_1\right)\left(c_1d_2-c_2d_1\right)e_3f_3\rangle 
=\frac{2}{n(n-1)(n-2)(n+2)}\times
\nonumber\\
& \times
\langle n(e\cdot f)\left[(a\cdot c)(b\cdot d)-(a\cdot d)(b\cdot c)\right]
\nonumber\\
& - (a\cdot c)\left[(b\cdot e)(d\cdot f)+(b\cdot f)(d\cdot e)\right]
  + (a\cdot d)\left[(b\cdot e)(c\cdot f)+(b\cdot f)(c\cdot e)\right]
\nonumber\\ \label{avabcdef2}
& + (b\cdot c)\left[(a\cdot e)(d\cdot f)+(a\cdot f)(d\cdot e)\right]
  - (b\cdot d)\left[(a\cdot e)(c\cdot f)+(a\cdot f)(c\cdot e)\right]
\rangle\,.
\end{align}

\end{appendix}

\end{document}